\documentclass[16pt,a4]{article}
\usepackage{epsfig}
\usepackage{setspace}

\title{White-Light Emission from Annealed ZnO:Si Nanocomposite Thin Films.}

\author{Shabnam$^a$, Chhaya Ravi Kant$^a$ and P.
Arun$^b$\footnote{email:arunp92@physics.du.ac.in, Telephone:091 011
29258401, Fax: 091 011 27666220} \\ \\
$^a$Department of Applied Sciences,\\ 
Indira Gandhi Institute of Technology,\\ 
Guru Gobind Singh Indraprastha University,\\ Delhi 110 006, India.\\
\\
$^b$Department of Physics \& Electronics,\\ S.G.T.B. Khalsa College,\\
University of Delhi, Delhi - 110 007, India\\
}

\begin{document}
\maketitle

\begin{abstract}
As grown ZnO:Si nanocomposites of different compositional ratios were fabricated by thermal evaporation techniques. These films were subjected to post deposition annealing under high vacuum at a temperature of $\rm 250C^o$ for 90min. The photoluminescence (PL) spectra of annealed samples have shown marked improvements both in terms of intensity and broadening. For the first time in ZnO:Si nanocomposite films we see huge UV, red and orange peaks at 310, 570 and 640nm. Structural and Raman analysis show formation of a Zn-Si-O shell around ZnO nano clusters wherein on heating $\rm Zn_2SiO_4$ compound forms. The new emissions are due to $\rm Zn_2SiO_4$ which completes white light spectrum. 
\end{abstract}

\vfil \eject

\section{Introduction}
Ever since the extraction of blue light from Mg doped Gallium Nitride(GaN) was made possible, research has been directed to yield cost effective white light emitting devices from a single chip. Owing to its structural similarity with GaN and wide band gap of $\rm \sim3.2eV$ ZnO is considered as a promising candidate for white-light production \cite{znoEg, ozgur}. In this direction, ZnO based nanostructures and nanocomposites have received ample attention. ZnO based nanocomposites have gained more importance in the field of white light emission. This is because an appropriate host material not only aids in broadening of light emission but also gives stabilty to the film by preventing agglomeration of the ZnO grains. To this effect, Silicon becomes a preferred choice to be used as host species \cite{siuse1, siuse2, 
siuse3}. This is because of the central role played by Silicon in the microelectronic industry and its known emisions in red region in its nanophase.
\par For obtaining cost-effective white light, Klason et al \cite{kal} have 
deposited n-type ZnO nanorods on p-Si. The films emitted white light on 
forward biasing. Depositing of ZnO nanoparticles in porous silicon (po-Si) 
has  also been reported by Bo et al \cite{z.bo} and Singh et al 
\cite{mehra}. Researchers like Pal \cite{pal} and Peng 
et al \cite{ypeng1} had studied the effect of varying ZnO to Silicon content obtained by RF sputtering. Preliminary investigations taken up by our group have reported discrete UV, blue, green and weak red wavelengths from ZnO:Si nanocomposites grown by vacuum thermal evaporation technique \cite{paper1, paper3}. It was also seen that an improvement took place 
in the luminescence emission of the samples by subjecting them to post 
deposition heat treatment \cite{paper2}. Post deposition treatments have been reported to bring variations in Photoluminescence spectra profile by other research groups as well \cite{pal, sha, znoheated}. In this manuscript we report the systematic study carried out to investigate the influence of annealing on ZnO:Si nanocomposites with varying ZnO:Si compositional ratio.

\section{Experimental}
\par ZnO:Si nanocomposites films were fabricated 
by thermally evaporating a mixture of powdered ZnO and n-Silicon at a vacuum of ${\rm \sim 10^{-6}}$ Torr in a Hind High Vac (Bengaluru), Thermal evaporation coating unit, Model 12A4D. The deposition was carried out on microcopic glass substrates maintained at 
room temperature. Starting material was 
prepared by mixing ZnO and Silicon in the proportions of 1:1, 1:2, 1:3, 2:3 
and 2:5 (by weight). These mixtures were then pelletized, to prevent its flying off the boats. 
As deposited nanocomposite films were subjected to annealing under high vacuum of the order of ${\rm \sim 10^{-6}}$ Torr and then allowed to cool naturally under vacuum. 
Films of varying composition with thickness 600\AA\ are named as sample (a1), (b1), (c1), (d1) and (e1) in order of the increasing Silicon content in starting material. The vacuum annealed counterparts are referred to as (a1v), (b1v), (c1v), (d1v) and (e1v).

\par The structural studies of the surface is measured by Pananalytical 
PW3050/60 Grazing Incidence angle X-Ray Diffractometer (GIXD) and that of the 
bulk region by Philips PW 3020 X-Ray Diffractometer (XRD). X-Ray 
Photoelectron Spectroscopy (XPS) was performed with Perkin-Elmer X-ray 
Photo-electron Spectrometer (Model 1257) with Al ${\rm K\alpha}$ (1486 eV) 
X-ray source. Photoluminescence (PL) scans were recored on Fluorolog Jobin Yvon spectroscope (Model 3-11) using an excitation wavelength of 270nm. Renishaw's ``Invia Reflex'' Raman spectroscope 
was used for measurements uisng $\rm Ar^{+2}$. The surface morphology and texture of the as 
grown nanocomposite films were studied using 
TECHNAI-20 $\rm G^2$ Transmission Electron Microscope (TEM). Below we
enlist the results of the various analysis done on our samples.

\section{Results and Discussion}
\subsection{X-Ray Diffracion \& Chemical Composition Studies}
\subsubsection{Structural Study}
The structural changes in the films caused by heat treatment was examined 
using GIXD. To compare the effect of heat-treatment, Figure 1(A) and 1(B) 
shows the GIXD scans of the samples after fabrication and post 
vacuum annealing. Similar to the pattern obtained for the as grown films, 
the GIXD scans of the annealed samples show peaks at 
$\rm 2\theta $=$\rm 36^0 $ and $\rm 43^0$. These peaks are the 
(101) and (102) plane of ZnO and Zn respectively.
A mixed response is seen on heat treatment with no variation in
samples (b1v) and (e1v), however, there is a marked increase in peak
intensities in (d1v) while it diminishes in (c1v). 

To investigate the increase in peak intensity of sample (d1v),
we have calculated the grain size of the pre and post annealed samples. The
grain size as calculated from the ZnO peak of (d1) was 8.7nm while that of sample (d1v) 
was found to be ${\rm 9.9 \pm 3nm}$. This marginal or no change in grain
size was evident in the other samples also. No growth in grain/cluster size is not surprising since the
ZnO nano-clusters are embedded in Silicon matrix which would deter any
further agglomeration. 

\subsubsection{Chemical Composition}
The surface and the film's bulk are usually different \cite{paper3} and
hence it is necessary to investigate the film below the surface. For this we
studied the samples using XPS not only at the surface but also beneath it by 
sputtering 125\AA\, of film layer. Figure 2(A) shows this depth profiling, 
recorded for Zn-$\rm 2p_{3/2}$ of sample (c1v). Presence of a single peak at 
around 1022eV suggests Zinc exists in the sample as ZnO. Contribution from 
elemental Zinc if any is insignificant. However, a shift to higher binding 
energy along the film thickness is visually evident from figure 2(A) (and
ploted in figure 3A). Before proceeding, it is worth mentioning that the 
existence of single Silicon peak in XPS as is the case with Oxygen
(figure 2B and 2C respectively) eliminates the possibility of 
existence of elemental Zinc and Silicon dioxide through out the film. 

\par In our previous report on as deposited films, we had observed that 
Zinc's ${\rm 2p_{3/2}}$ peak shift towards lower energy side along the film
thickness. We believe this shift is related to the change in the 
neighboring environment in terms of relative ZnO to Silicon abundance.
However, here we see a shift in Zinc's peak towards the higher energy side
as we move deeper into the film.

\par Following the methodology we adopted in our earlier work \cite{paper3}, 
we determined the ratio of ZnO to Si along the thickness of the film. Figure 
3(B) shows the linear decrease of this ratio with film depth. Decrease in 
ZnO:Si ratio from 1.023 to 0.271 along the depth of the film can also be 
interpreted as an increase in the Silicon content with the depth of the film. 
Combining the results of Figure 3(A) and 3(B), figure 3(C) shows the variation of 
ZnO peak position with ZnO:Si ratio. The data indicates a shift in Zinc
${\rm 2p_{3/2}}$ peak to the lower energy side with decreasing Silicon
environment. Though the samples under study here and in our previous study
were different, the fact that Zinc ${\rm 2p_{3/2}}$ peak moves to the lower 
energy side with decreasing Silicon environment is consistent. Thus,
structurally, morphologically and compositionally vacuum annealing has not
effected our sample.

\subsection{Raman Spectra}
We have noticed that Raman spectra reveals more information on the structural
properties of our nano-composites than X-Ray diffraction \cite{paper3}.
Hence, to investigate the structural modification incorporated in our
samples on vacuum annealing we analysed the samples using Raman 
spectroscopy. The Raman spectra was taken in standard back scattering 
geometry using Argon ion laser for excitation. Figure 4 compares the Raman 
spectra of the as grown samples (a1-e1) with their vacuum annealed 
counterparts (a1v-e1v). The as grown films irrespective of the ZnO/Si
starting ratio gave broad spectra in 300-600$\rm {cm^{-1}}$ range. The 
spectras were essentially three prominent unresolved peaks, namely at 
$\rm \sim 310, 440$ and 565 $\rm {cm^{-1}}$. The $\rm \sim 310 cm^{-1}$ peak 
corresponds to the LA mode of amorphous Silicon \cite{paper1, paper3}.
Similarly, the ${\rm \sim 440cm^{-1}}$ and ${\rm \sim 565cm^{-1}}$ peaks are 
attributed to Wurtzite ZnO bond vibrations and defect related bondings in
ZnO respectively \cite{raman1, raman2}.

Visual examination show significant changes not only in peak sizes but also
in their positions. For example, the plot of figure 5(A) shows variation in 
ratio of area enclosed by the ${\rm 440cm^{-1}}$ peak of the vacuum annealed 
sample to that of as grown with ZnO content. A decreasing trend is seen with
increasing ZnO content. We believe that this decrease manifests due to 
decrease in ordering in the ZnO. Also, the ordering is easily broken in
samples with higher ZnO content on vacuum annealing. Even the peak position 
(440$\rm {cm^{-1}}$) of the annealed samples 
as compared to those of as deposited peaks show a linear trend with the peak 
moving to a higher wave-number on annealing (fig 5B).
That is, on vacuum annealing the peak position corresponding to bondings 																																																																																																																																																																																																																			of the Wurtzite structured ZnO shows increased wave-number where the
increase is more substantial if the as grown samples peak was at a higher
wave number. This too must be indicative of increased disorder. Also, 
Yadav et al
\cite{vinaygupta} have reported an increase in wavenumber with decreasing
grain size. This would reason that vacuum annealing has resulted in a
decrease in grain size (except for sample a1v).

\par This lack of ordering discussed above should reflect in the vacuum
annealed sample's $\rm \sim 565 {cm^{-1}}$ peak that corresponds to the 
defects in ZnO \cite{zniraman}. Plots of ${\rm \Delta_{VA}/\Delta_{AG}}$ (${\rm
\Delta}$ represents peak area of vacuum annealed `VA', and as grown `AG')
and ${\rm I_{VA}/I_{AG}}$ (peak intensities) with varying ZnO content shows
linear trend. Figure 5(C) shows the variation in ratio of intensities with ZnO
content. As expected, the linear trend with positive slope suggests an 
increase in the defects in vacuum annealed ZnO:Si nanocomposites for samples
with large ZnO content. There also data point of `d1v' stands apart from
the trendline. However, this indicates `d1v' contains appreciable defects
along with substantial ordering. The fact that sample `d1v' shows a 
simultneous increase in
crystallinity and defects appears contradictory at first glance, however, we
have been able to show that a sample with comparable amounts of defects
co-existing with good ordering gives broadening in PL spectra \cite{paper3}. 
Figure 6 gives
a plot of relative presence of Wurtzite ZnO to ZnO with defects ($\rm \Delta 
438cm^{-1}/ \Delta 565cm^{-1}$) with varying ZnO content. This
graph helps in predicting broadening of PL spectra. 
In an earlier work on as grown samples (shown by filled circles in
figure 7), it was observed that samples which had nearly equal areas (and
hence ratio $\sim$1) showed maximum broadening in PL \cite{paper3}. It can
be appreciated from the plot that the sample `d1v' has nearly equal areas of
the peaks resulting from defect free and defect related ZnO lattices. 
Moreover, sample `d1v' lies on the minimum of the curve (visual aid showing 
the trend), so it is expected to show maximum broadening in photoluminescence. 
The downward shift of the `a1v' data point shows that this sample has also
got comparable contributions from wurtzite and defect related ZnO structures.
As per our prediction, `a1v' should also show broadening in
PL spectra along with the other heat treated samples.

Finally we now comment on the reduction in the $\rm \sim 310 cm^{-1}$ peak 
of annealed samples as compared to that in the as deposited samples. In our 
ZnO:Si nanocomposites, we have ascribed this peak to LA mode of amorphous 
Silicon \cite{paper1, paper3}. Since the annealing of the samples has been 
done in the high vacuum of the order of $\rm 10^{-6}$ torr, as also by our
XPS results we rule out the possibility of Silicon's oxidation. Hence, we
believe this reduction is due to some improvement in ordering of Silicon.
However, since XRD failed to give a nanocrystalline peak of Silicon, we 
believe only short-range ordering of amorphous silicon has taken place.

\subsection{Photoluminescence Spectra}
In our previous works we have shown that a broad spectra is achievable from ZnO:Si nanocomposite films. However, they had poor or no emission in the red wavelength region. To obtain further broadening and to improve emission in the red wavelength region, we annealed our samples under high vacuum. 
Figure 7 compares the PL spectra of samples (a1) and (a1v). Broadening accompanied with a multifold increase in intensity on annealing can be easily appreciated. In the absence of prominent shoulders resulting from broadening of peaks we were not able to deconvolute the spectra using standard softwares. However, based on our knowledge from study of asgrown samples we expect peaks at 365 and 420nm. These peaks correspond to band edge emission from Wurtzite ZnO and the interfacial layer between ZnO grains and Si background. These two peaks in the annealed samples were unresolved and hence using Peakfit-4 we have placed a peak at 395nm (fig 8). 
In the previous section, Raman analysis suggested a decrease in the grain size of ZnO accompanied with reduction in ${\rm 438cm^{-1}}$ peak implying decrease in Wurtzite ordering. A direct relationship between the Raman ${\rm 438cm^{-1}}$ peak and PL's 365nm was established \cite{paper3}. Thus one expects a reduced contribution from the band edge emission peak of ZnO. However, since the unresolved peak remains significant, one can infer an improvement in contribution from the interface, i.e. the 420nm peak. Another peak of the blue region which appears in asgrown samples and presists even after annealing is the 470nm peak. We attribute this peak too to the interface \cite{ypeng1, paper3}.

\par Increasing ZnO content in this study showed decreasing grain size, hence the 430nm peak seems to be enhanced for samples with smaller grain size. A question that would need answering would be {\sl ``how can diminishing grain size contribute more to an emission process?"}. To analyse this we took a sample (this sample was used in our earlier study \cite{paper2}) and heated it at ${\rm 250^oC}$ for intervals of 30, 60, 120min. The grain size was found to decrease with increased heating. We had proposed the formation of a ``shell" around the increasingly crystalline ZnO with Zn-Si-O shell material growing into the asgrown cluster, thus reducing the grain size. In fig~8 we plot the PL 430nm peak with increasing shell volume. Shell volume $\rm V_{shell} $ is 
\begin{eqnarray}
V_{shell} \propto R^3_o-R^3_n\nonumber
\end{eqnarray}
where, $R_o$ is the grain size of the as grown sample 
and $R_n$ is that of annealed samples. 
We find a linear co-relation, suggesting that the 430nm PL emission peak intensity is due to the ``shell" volume or in turn amount of Zn-Si-O linkages present.

The PL spectra was recorded without use of optical filters. This results in the presence of a huge peak at 540nm which is called the second harmonic peak. As compared to the unheated samples PL spectra, the neck of this harmonics is quite broad (example fig~6). This broadening maybe due to existence of the 540nm green peak associated with oxygen vacancy in ZnO. While this peak was resolvable in as grown sample's PL sectra from the second harmonic peak, we were not able to do the same in this study.

\par The other two peaks lying at ${\rm 570nm}$ and ${\rm 640nm}$ with marked contributions have been observed for the first time in our samples. While ${\rm 570nm}$ peak could be ascribed to emission between energy levels caused by Zn interstial defects \cite{zniyellow} it does not explain the existence of the ${\rm 640nm}$ peak. Contribution of emission from nano-Si can be ruled out as per our XRD and Raman analysis. We believe vacuum annealing has transformed some of the ZnO and Si linkages present in the shell into 
$\rm Zn_2SiO_4$. $\rm Zn_2SiO_4$ is known to emit strongly in red, orange and UV regions \cite{willemite1, willemite2}. In fact one can notice the UV emission at 310nm is strong whenever an intense 570 and 640nm emission peak exists. We noticed that the sample (c1v), which has significant red and orange emissions also show a huge 310nm peak.

\section*{Discussion}
 Based on our previous studies and results from the present work, we now are in a position to explain the processes taking place in ZnO:Si nanocomposite films. While the Silicon matrix prevents ZnO grains from agglomerating the unsaturated bonds of ZnO and Si at the interface form Zn-Si-O linkages. The interface thus contributes to emissions at 420nm and 470nm. Due to the thermal evaporation technique used for fabriaction, ZnO clusters have defects. While the ordered bondings give rise to emissions at 350nm, those associated with defects give emissions at 540nm. based on relative proportion of the two, the contribution of emission varies. We have shown comparable proportions give comparable emissions and has a broadening of PL in the blue-green region. Annealing results in migration of defects from the core of ZnO to the interface. This results in increased Zn-Si-O bondings that go on forming a shell which expands inwards decreasing ZnO grain size. Appropriate temperature of annealing leads to the formation of 
$\rm Zn_2SiO_4$ within the shell that leads to new emissions in UV and red regions. As stated the present work along with our initial results give a good insight of the material process taking place and gives scope of understanding how to engineer samples for broad emission white light LEDs based on ZnO:Si nanocomposites.

\section*{Conclusion}
As deposited nanocomposites of ZnO:Si of various compositional ratios were annealed under high vacuum. Structural and compositional studies carried out by X-Ray Diffraction and XPS were not able to detect appreciable variation. However, PL spectra have shown a significant improvement. Not only have the intensities increased but appearance of three new peaks at 310, 570 and 620nm can be easily noticed. These peaks appeared along with earlier reported peaks in the blue-green region. Annealing has not only pushed the interface into the ZnO grain but also transformed Zn-Si-O linkages present at interface into  $\rm {Zn_2SiO_4} $. We attribute the new peaks to $\rm {Zn_2SiO_4} $ existing in the shell surrounding the ZnO grains. This study further substantiates our earlier claims correlating broadening in blue-green region to existence of appropiate ratio of two phase, namely crystalline and defect related ZnO. Sample (a1v) shows improved broadening as compared to its as grown counterparts, due to appropriate mix of the two phases achieved after annealing in vacuum. Presence of all the wavelengths completes our white-light spectrum. We thus successfully show that suitable ZnO:Si compositional ratio with appropriate post deposition treatment is crucial for obtaining white light. This study along with more careful investigations should pave the way for future white light emitting devices

\section*{Acknowledgment}
 We are thankful to Dr.D.K.Pandaya at Indian Institute of Technology, Delhi for GIXD measurements. The resources utilized at  
University Information Resource Center, Guru Gobind Singh Indraprasta
University is gratefully acknowledged. We also would like to express our gratitude to Dr. Kamla Sanan and Dr. Mahesh Sharma (both at National Physical Lab., Delhi) for carrying out the photoluminescence and XPS studies respectively. Author CRK is thankful to University Grants Commission 
(India) for 
financial assistance in terms of research award, F.No-33-27/2007(SR). One of the authors PA is grateful to Department of Science and Technology (India) for funding present work with research project SR/NM/NS-28/2010.                     . 

\newpage
\section*{Figure Captions}
\begin{itemize}
\item[1.] GIXD scans of (A) as grown and (B) vacuum annealed samples.
\item[2.] XPS depth profile scans of (A) $\rm 2p_{3/2}$ peaks of Zinc (B) Silicon and (C) Oxygen of sample (c1v).
\item[3.] (A) Variation of Peak position of Zn ${\rm 2p_{3/2}}$ with 
depth (B) Fraction of Zinc in bonding to amount of Silicon present along the 
thickness and (C) Peak position of Zinc in bonding with Oxygen to its fraction of presence.
\item[4.] Raman spectra of (A) as grown samples (a1), (b1), (c1), (d1), (e1) and (B) vacuum annealed samples (a1v), (b1v), (c1v), (d1v) and (e1v). Also seen are the deconvoluted peaks assigned to amorphous silicon ($\rm 310cm^{-1}$), wutzite structure of ZnO ($\rm 438cm^{-1}$) and with defect related peak of ZnO ($\rm 570cm^{-1}$).
\item[5.] (A) Relative change in Intensity of 438 $cm^{-1}$ peak in vacuum annealed samples to that in as grown samples with respect to zno content, (B) Variation of peak position in the 438 $cm^{-1}$ peak in vacuum annealed samples with respect to that in the as grown samples and (C) Relative change in Intensity of 560 $cm^{-1}$ peak in vacuum annealed samples to that in as grown samples with respect to ZnO content.
\item[6.] Relative presence of defedt related ZnO to wurtzite ZnO (Area 438$cm^{-1}$/Area 565$cm^{-1}$ from Raman Spectra) for varying ZnO content in film for (A) vacuum annealed (B)as deposited films.
\item[7.] PL spectra of samples (a1) and (a1v). (Counts of (a1) have been scaled by 3 (i.e.X3) to compare the spectra.)
\item[8.] PL of samples (a1v), (b1v), (c1v) and (d1v).
\item[9.] Plot of shell volume with respect to intensity of 430nm observed in photoluminescence (method described in the text).

\end{itemize}

\newpage

\vfil \eject
\pagestyle{empty}
\newpage

\begin{figure}[h]
\begin{center}
\epsfig{file=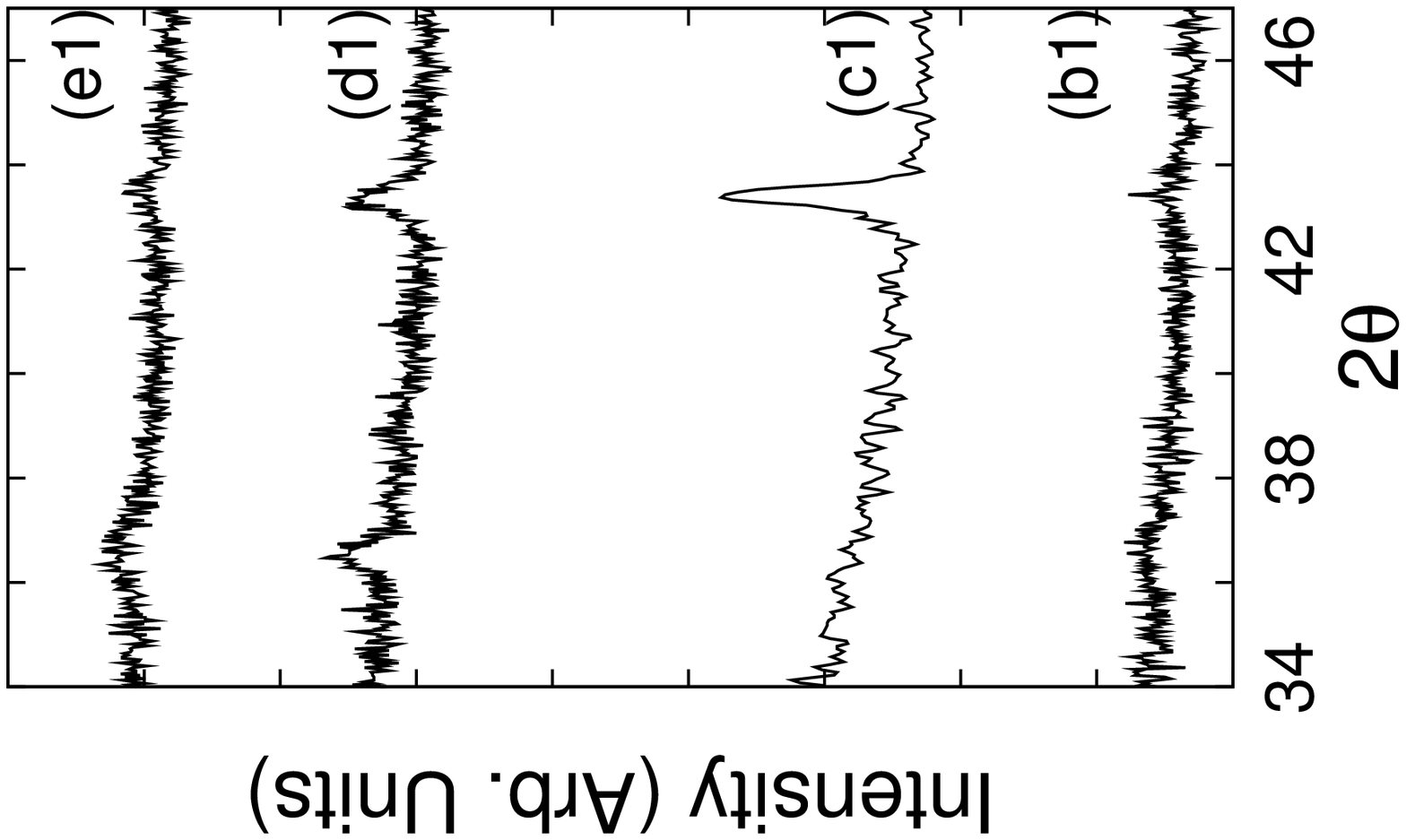, width=3in, angle=-90}
\hfil
\epsfig{file=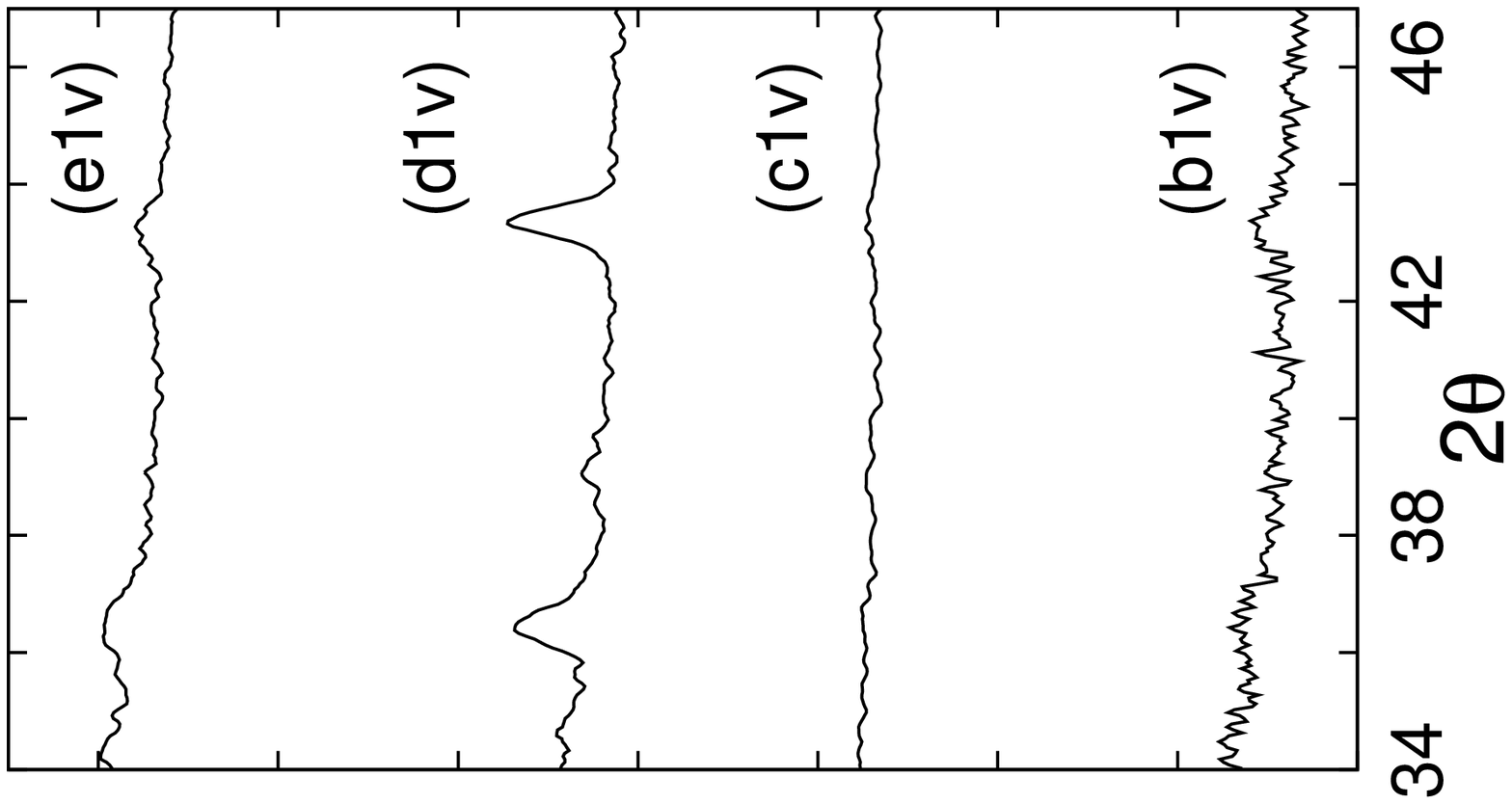, width=3in, angle=-90}
\end{center}
\caption{\sl GIXD scans of (A) as grown and (B) vacuum annealed samples.}
\end{figure}

\begin{figure}[h]
\begin{center}
\epsfig{file=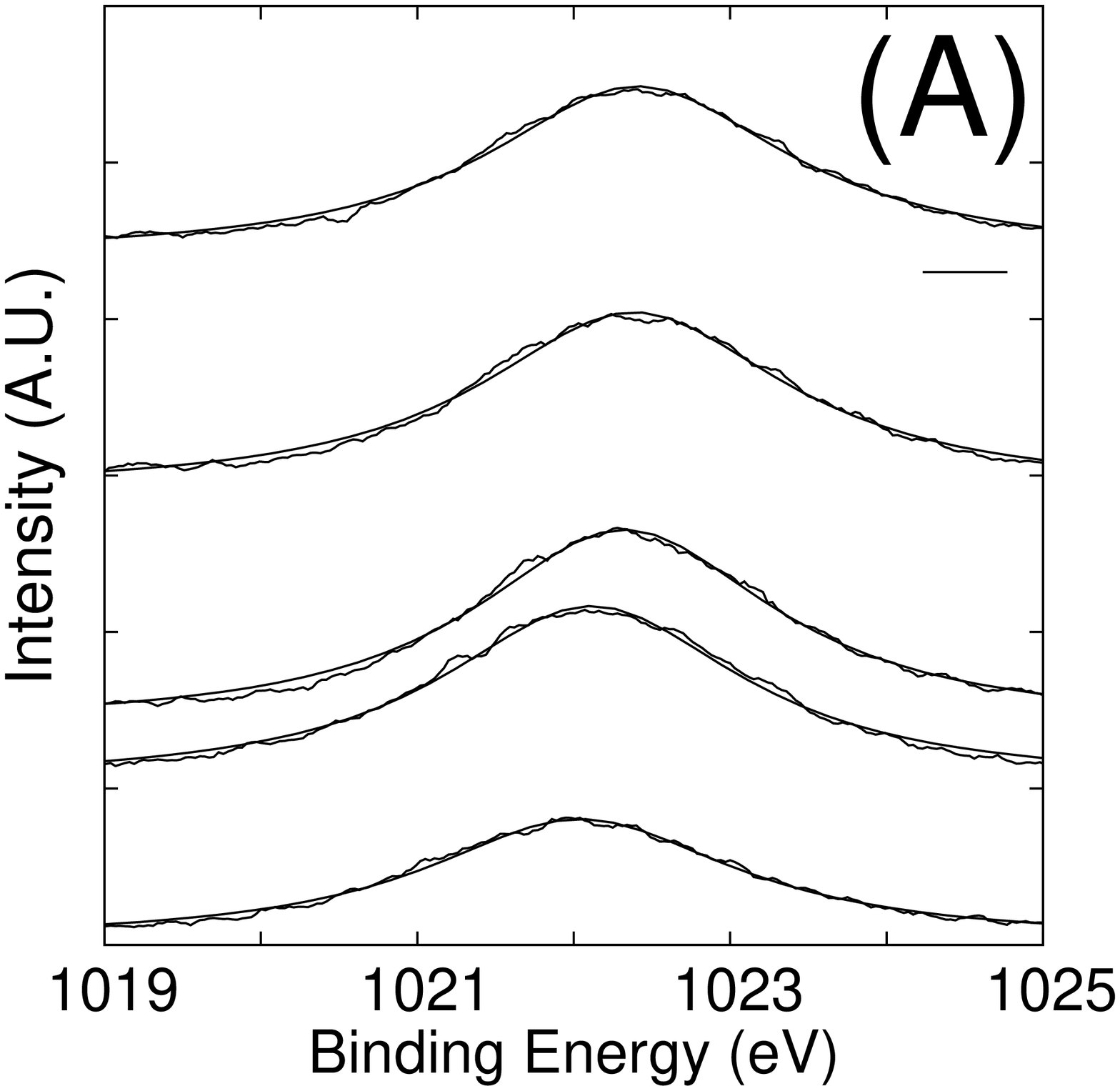, width=1.5in, angle=-90}
\hfil
\epsfig{file=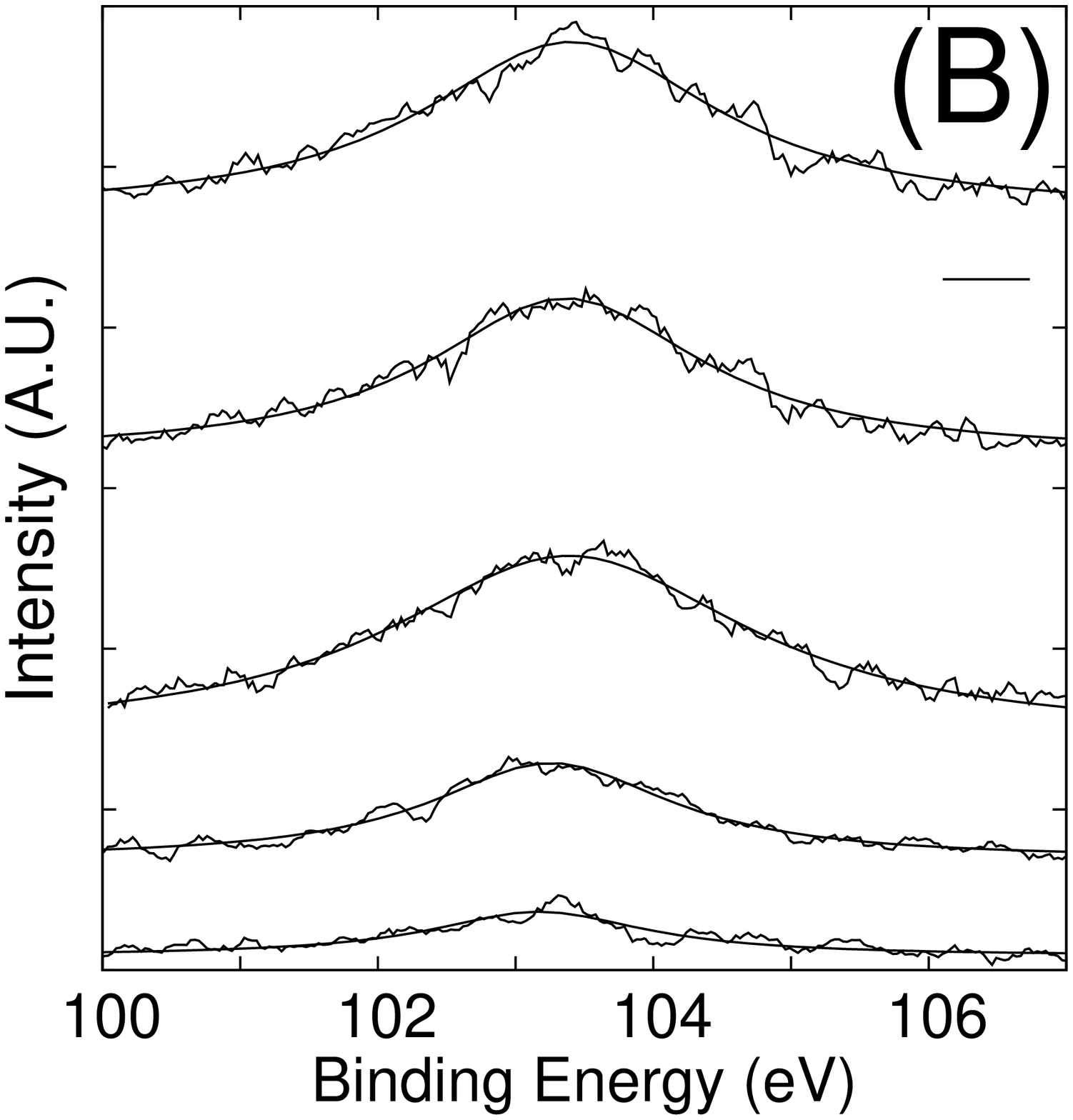, width=1.5in, angle=-90}
\hfil
\epsfig{file=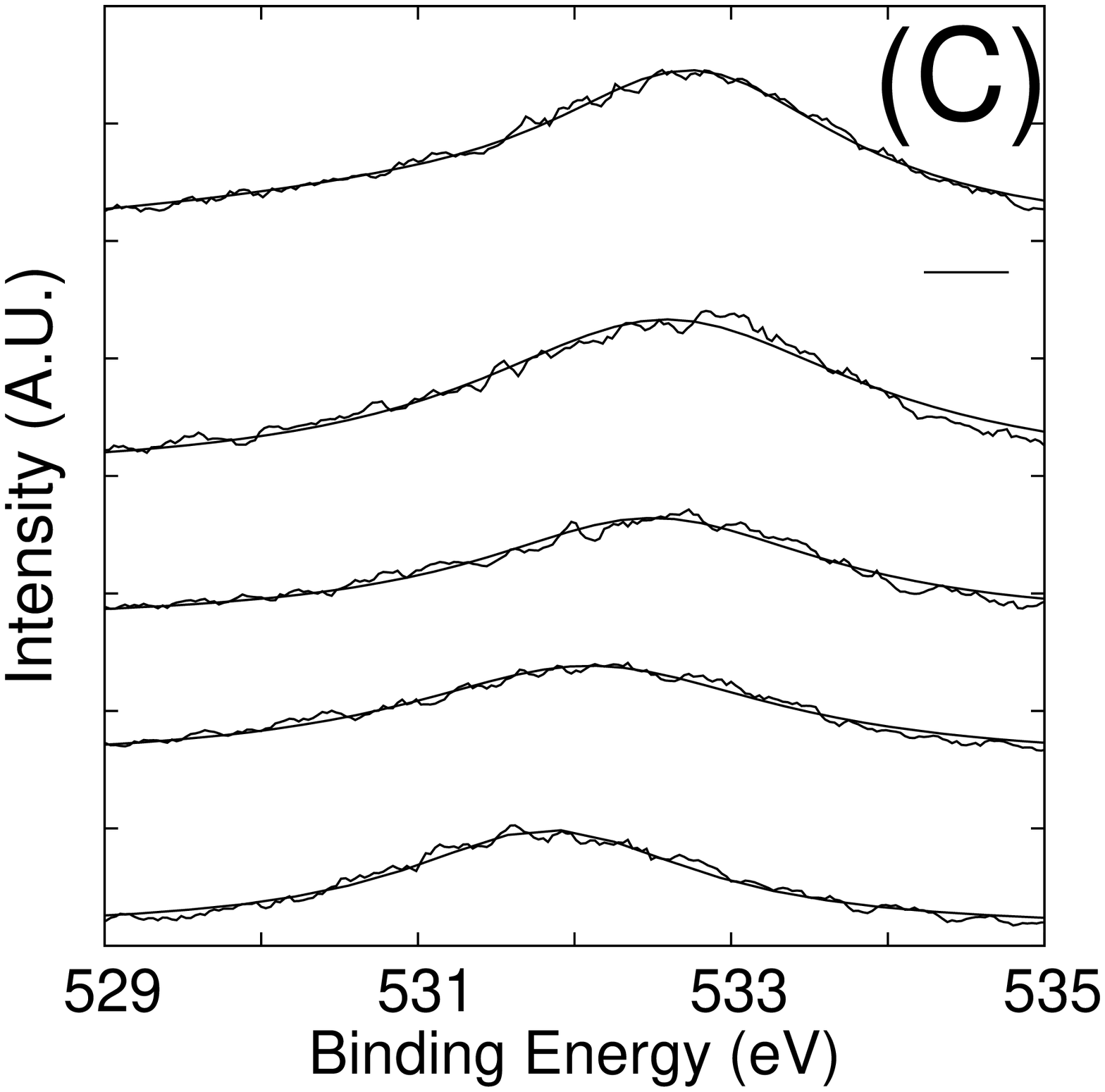, width=1.5in, angle=-90}
\end{center}
\caption{\sl XPS depth profile scans of (A) $\rm 2p_{3/2}$ peaks of Zinc (B) Silicon and (C) Oxygen of sample (c1v).}
\end{figure}

\begin{figure}[h]
\begin{center}
\epsfig{file=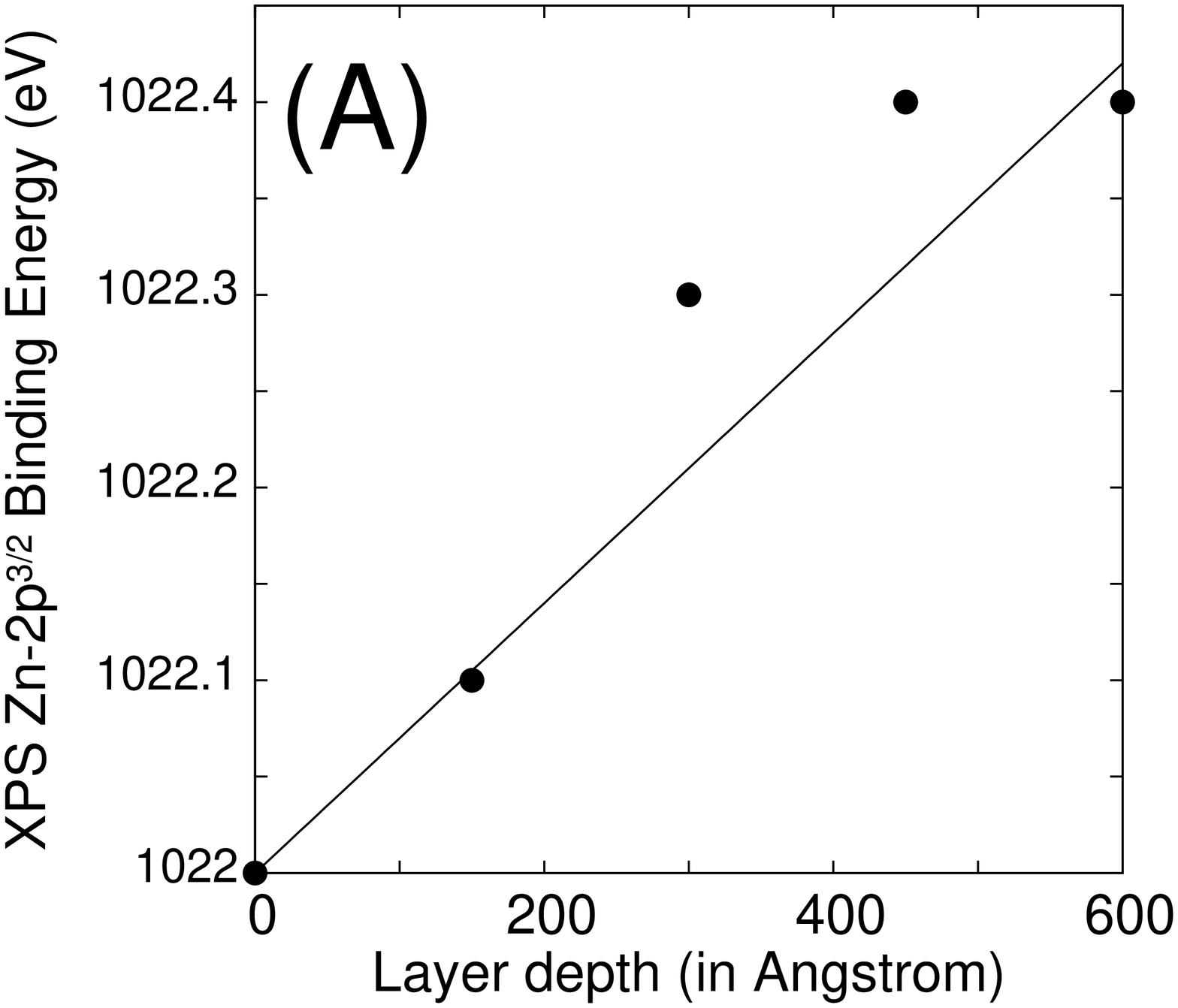, width=1.5in, angle=-90}
\hfil
\epsfig{file=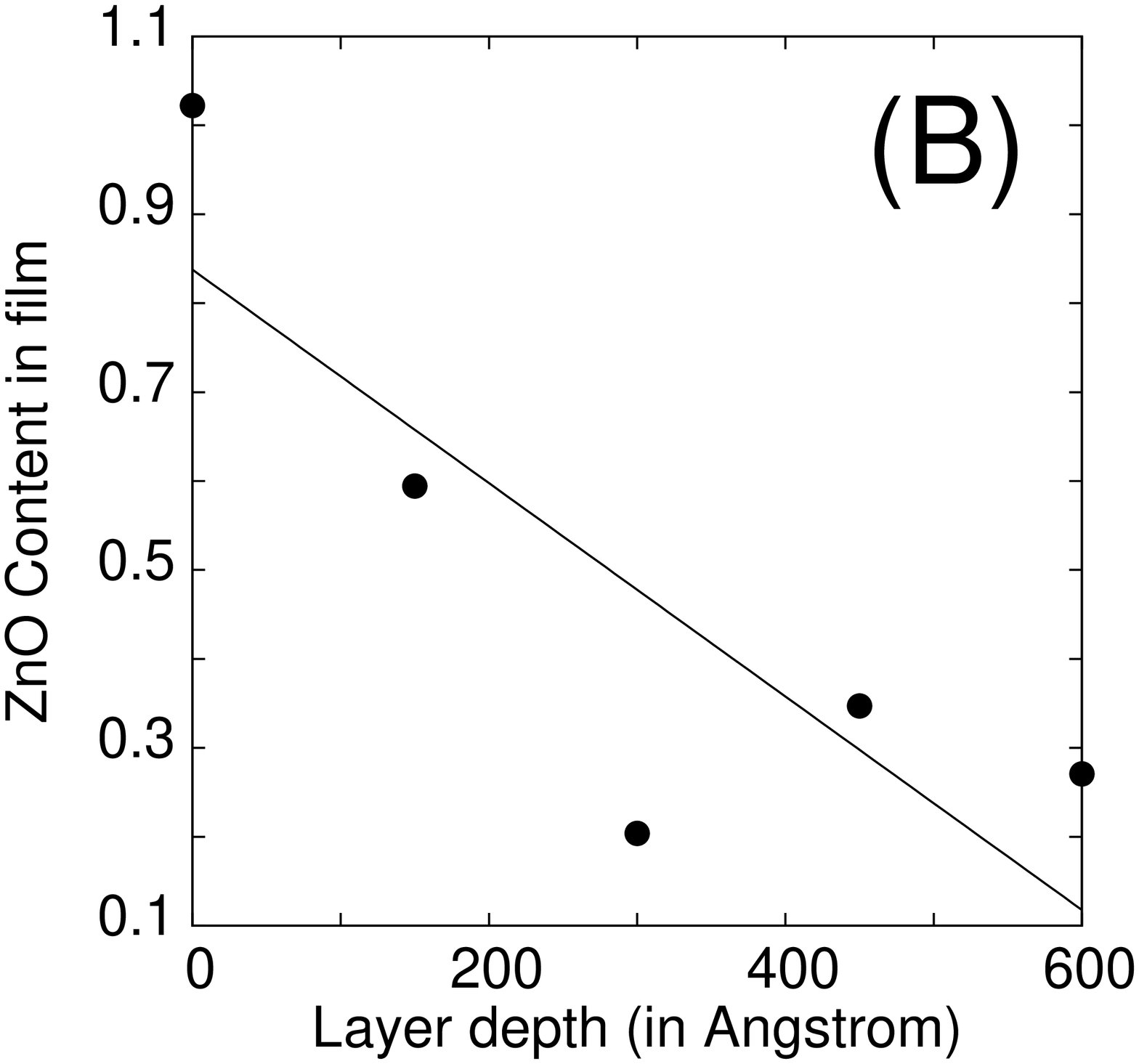, width=1.5in, angle=-90}
\hfil
\epsfig{file=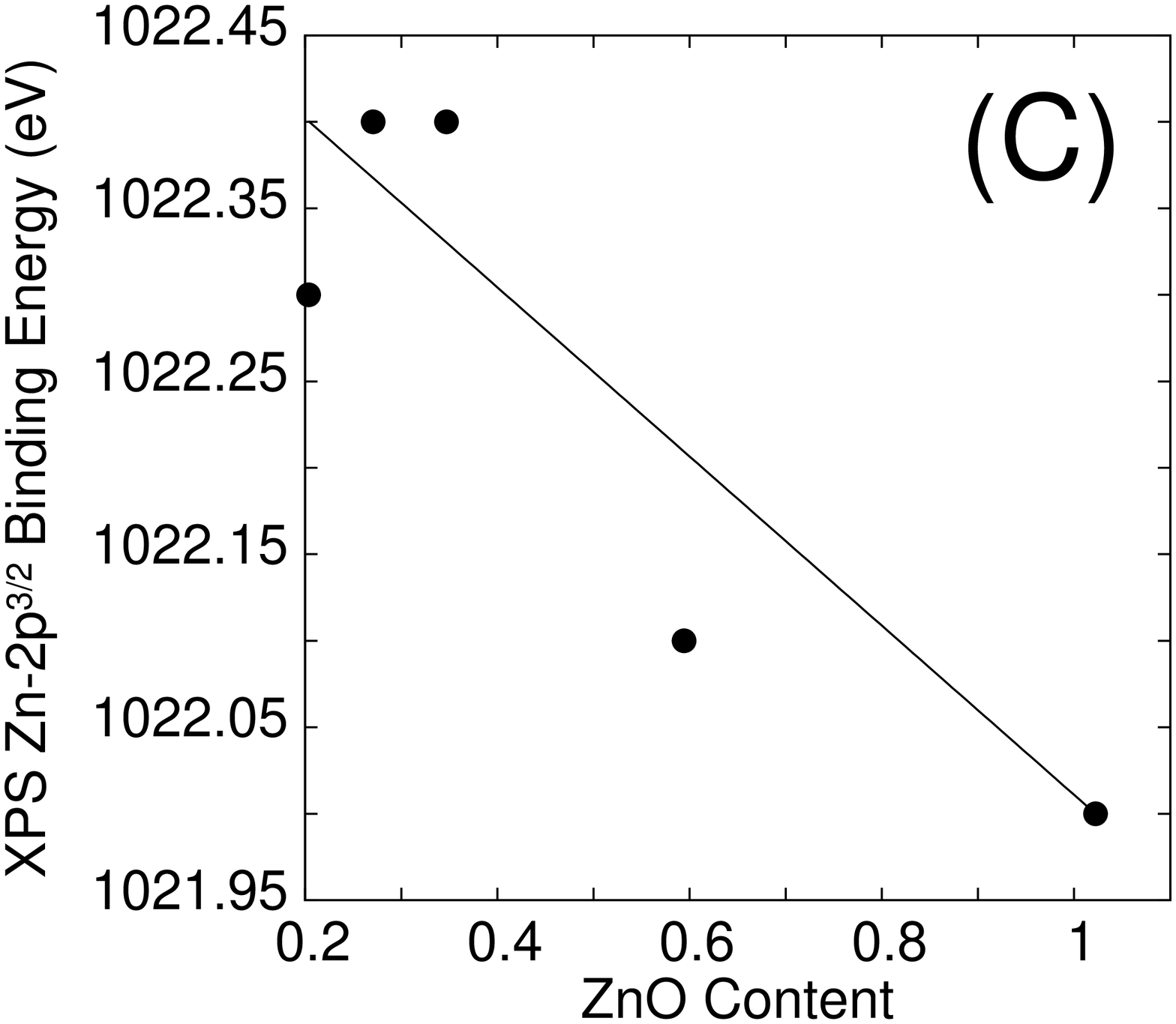, width=1.5in, angle=-90}
\end{center}
\caption{\sl (A) Variation of Peak position of Zn ${\rm 2p_{3/2}}$ with 
depth (B) Fraction of Zinc in bonding to amount of Silicon present along the 
thickness and (C) Peak position of Zinc in bonding with Oxygen to its fraction 
of presence. }
\end{figure}

\begin{figure}[h]
\begin{center}
\epsfig{file=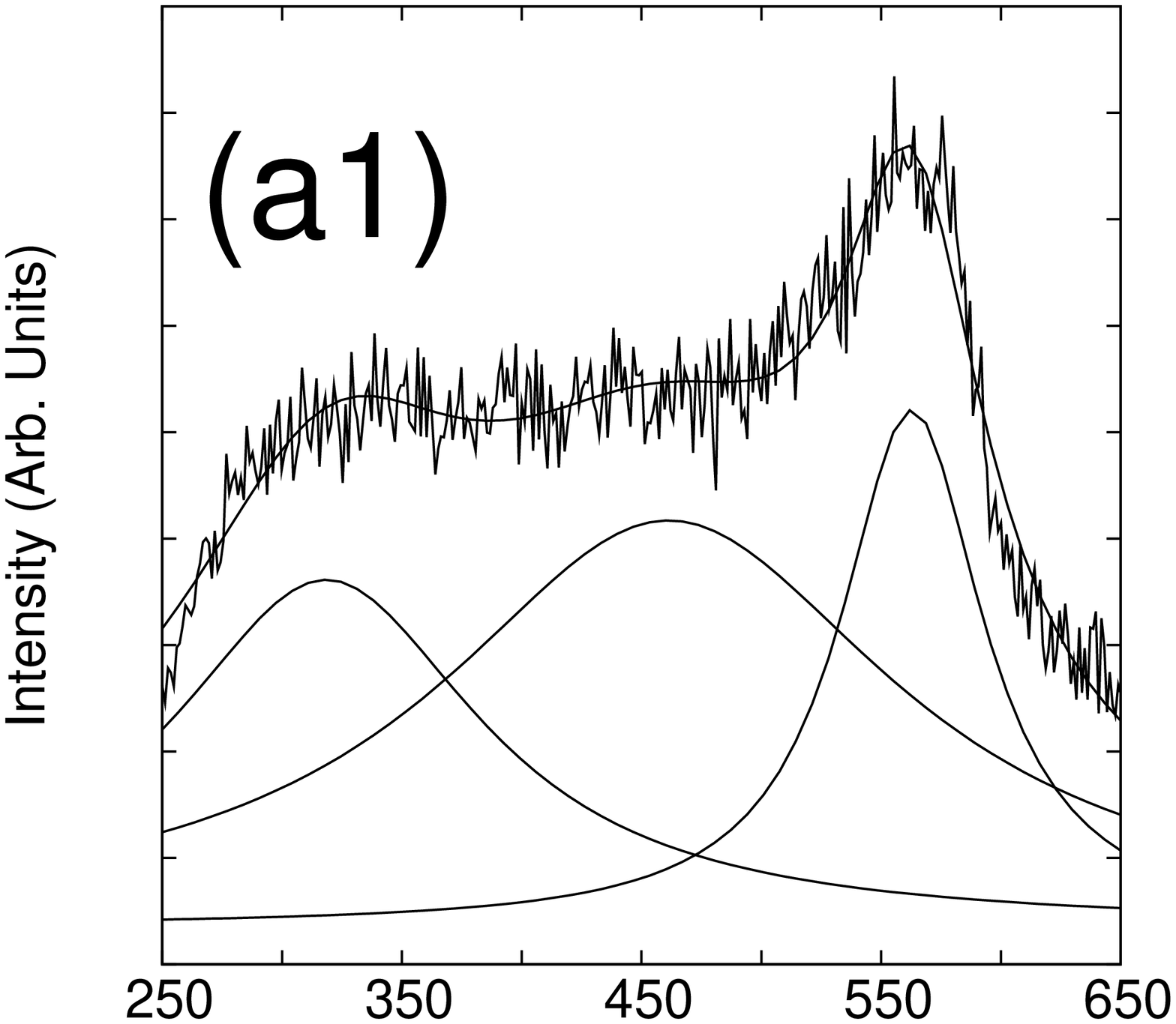, width=0.9in, angle=-90}
\hfil
\epsfig{file=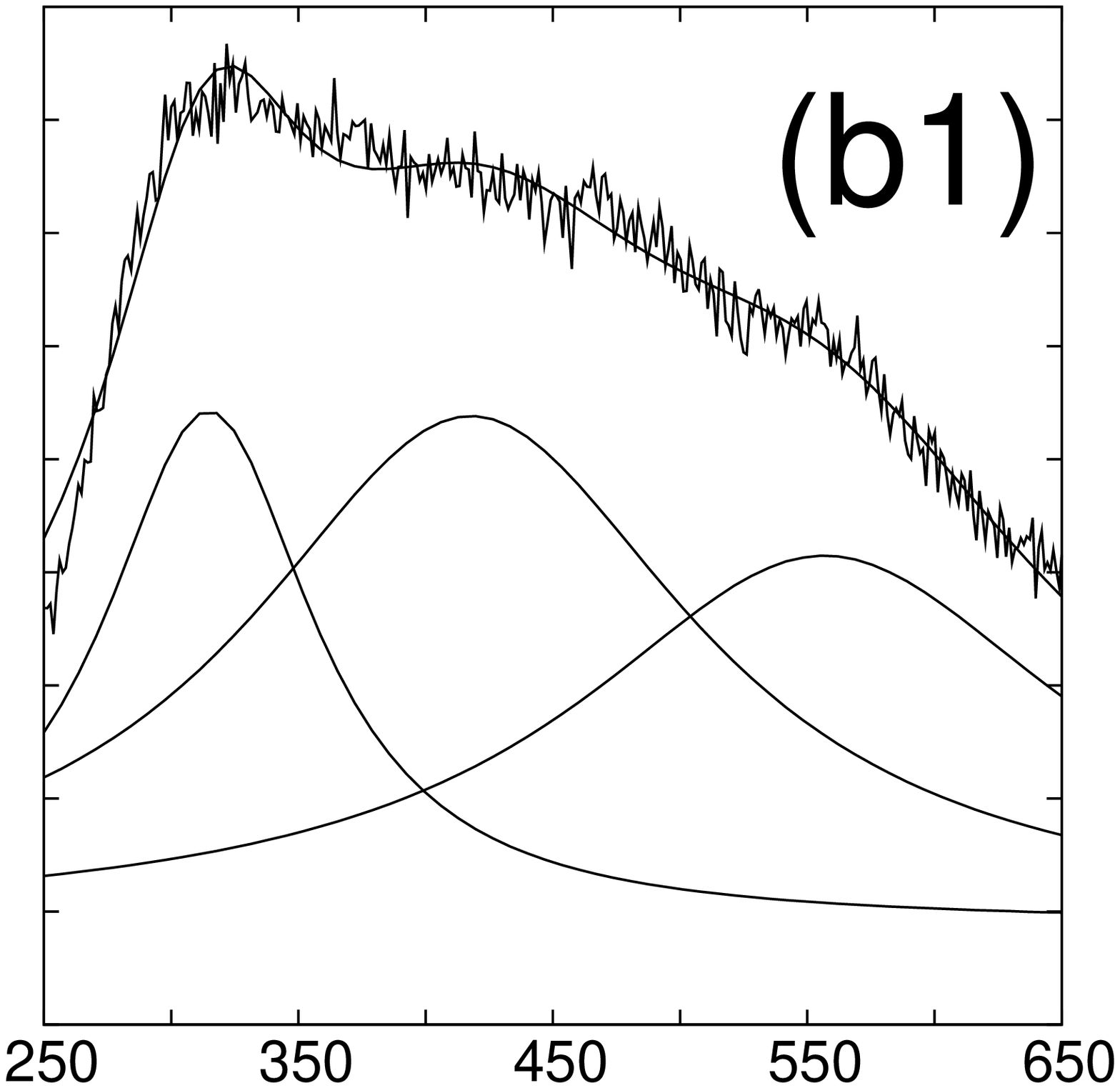, width=0.9in, angle=-90}
\hfil
\epsfig{file=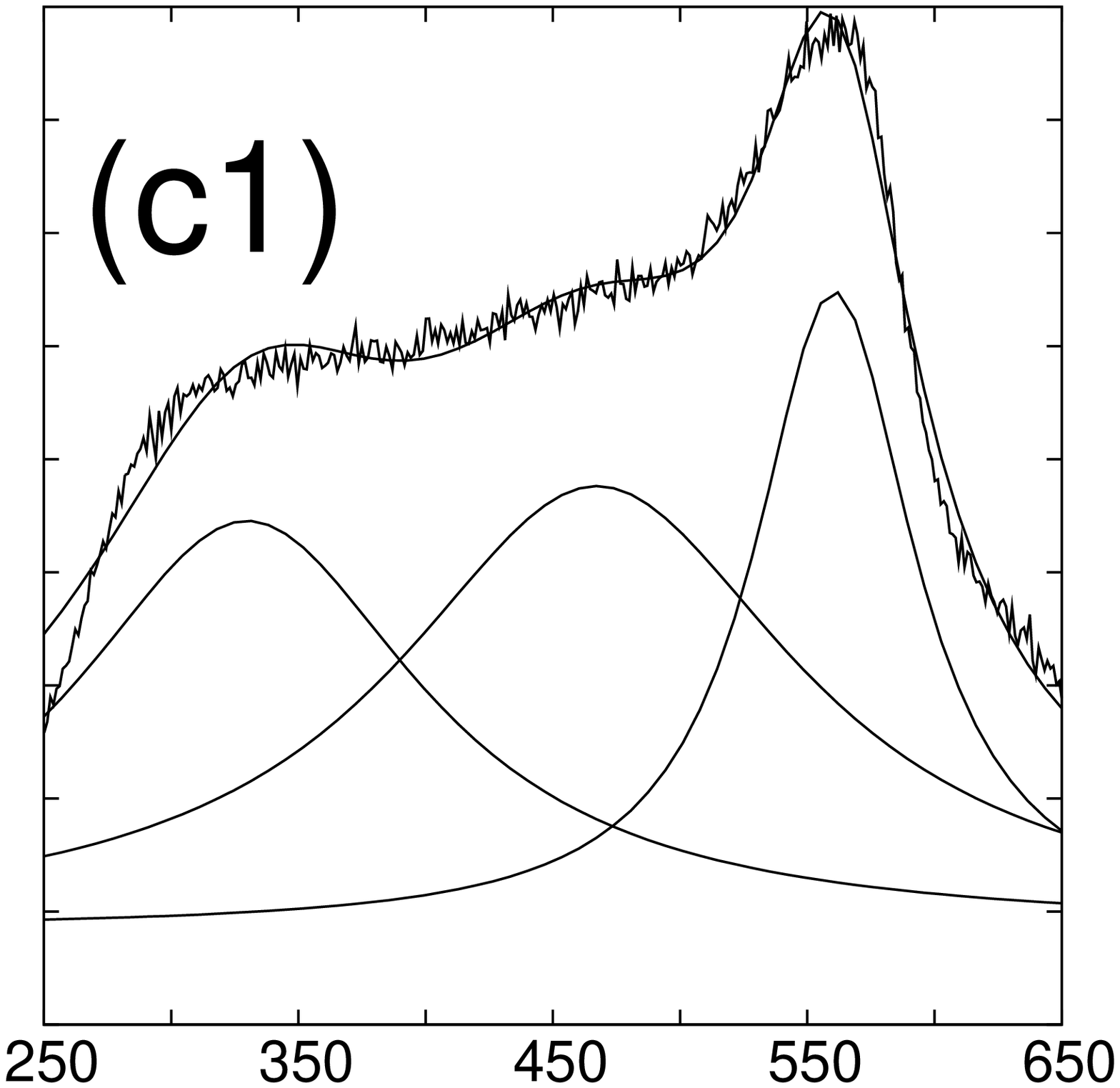, width=0.9in, angle=-90}
\hfil
\epsfig{file=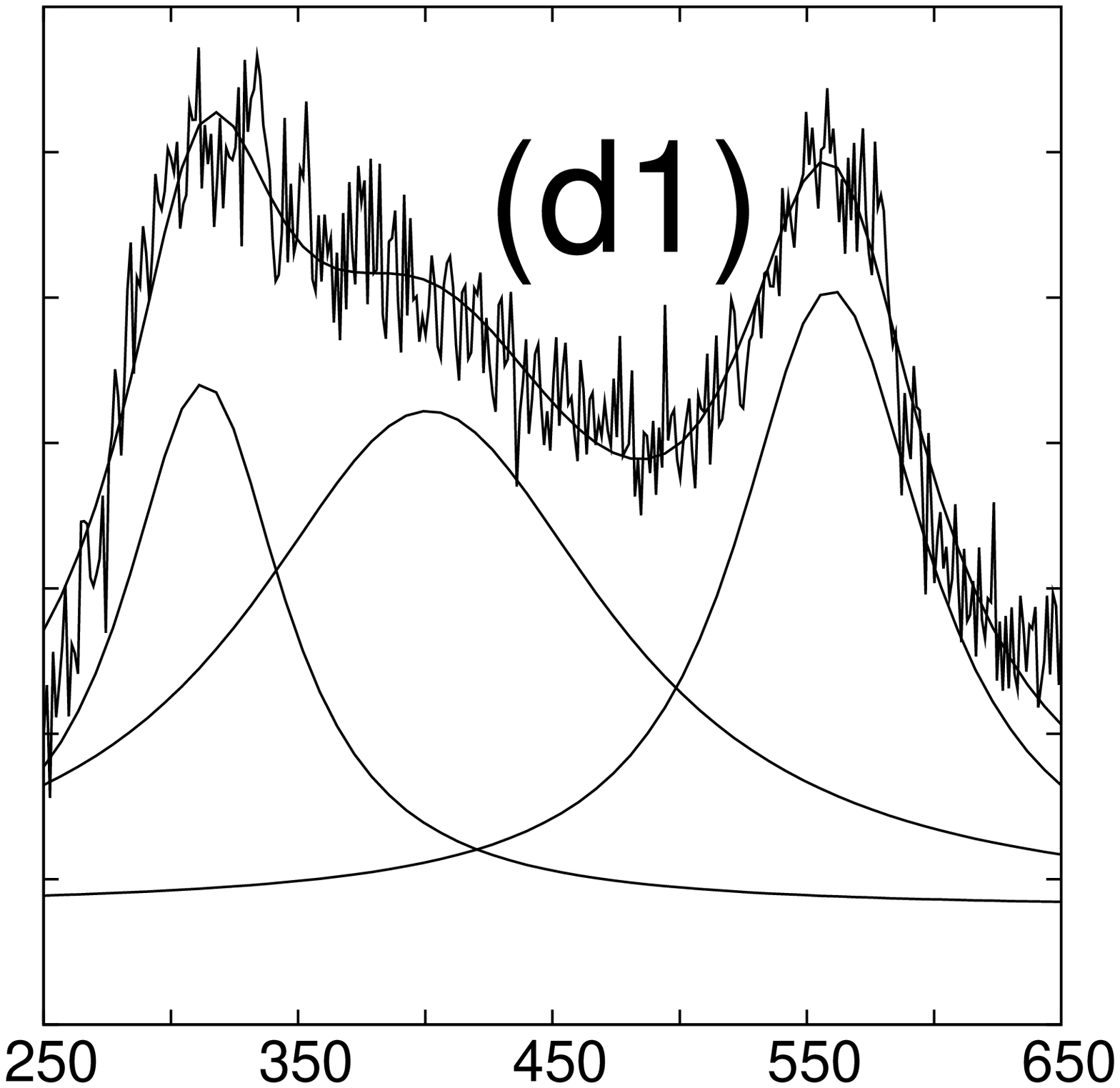, width=0.9in, angle=-90}
\hfil
\epsfig{file=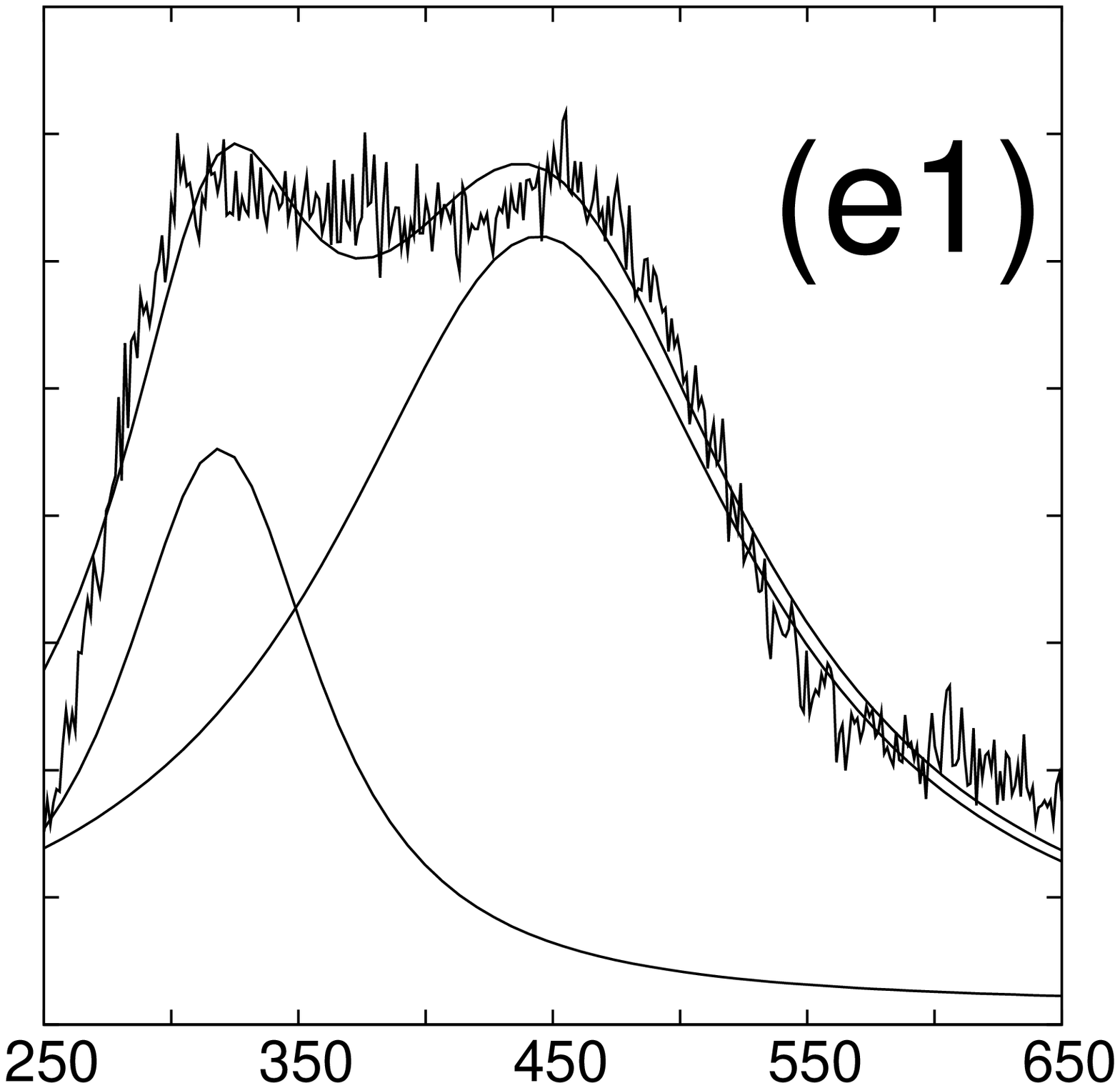, width=0.9in, angle=-90}
\vfil
\epsfig{file=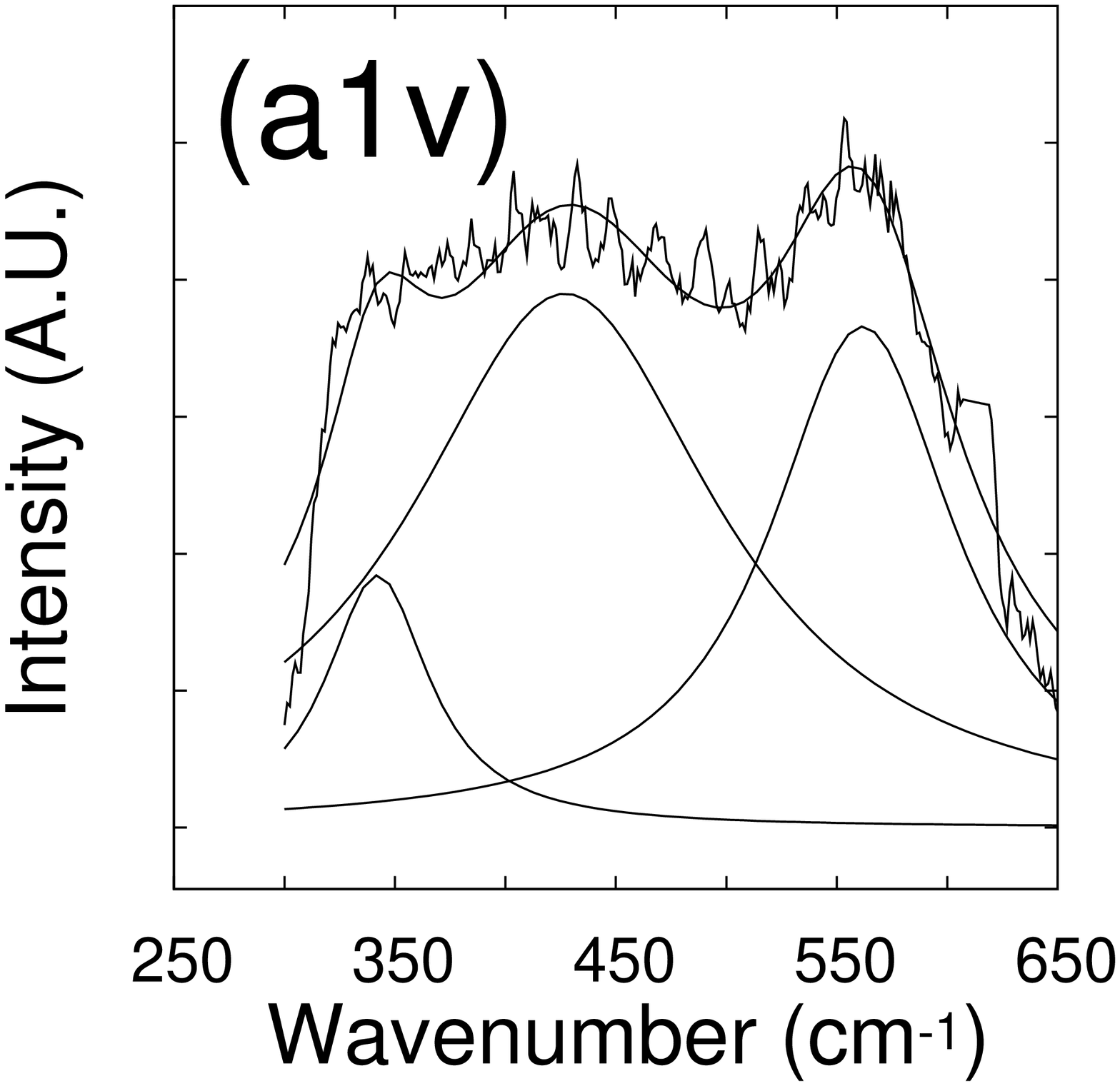, width=0.9in, angle=-90}
\hfil
\epsfig{file=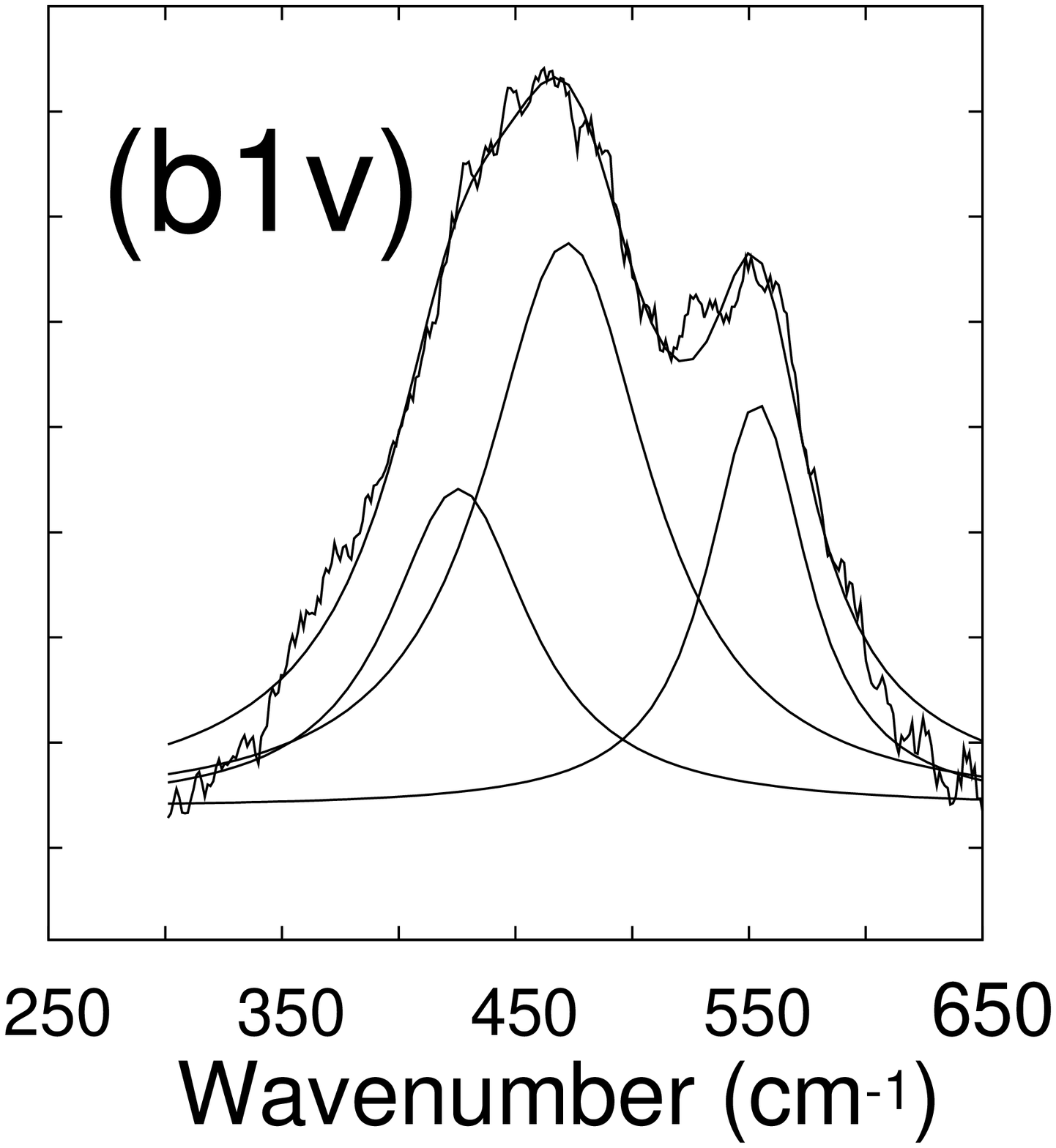, width=0.9in, angle=-90}
\hfil
\epsfig{file=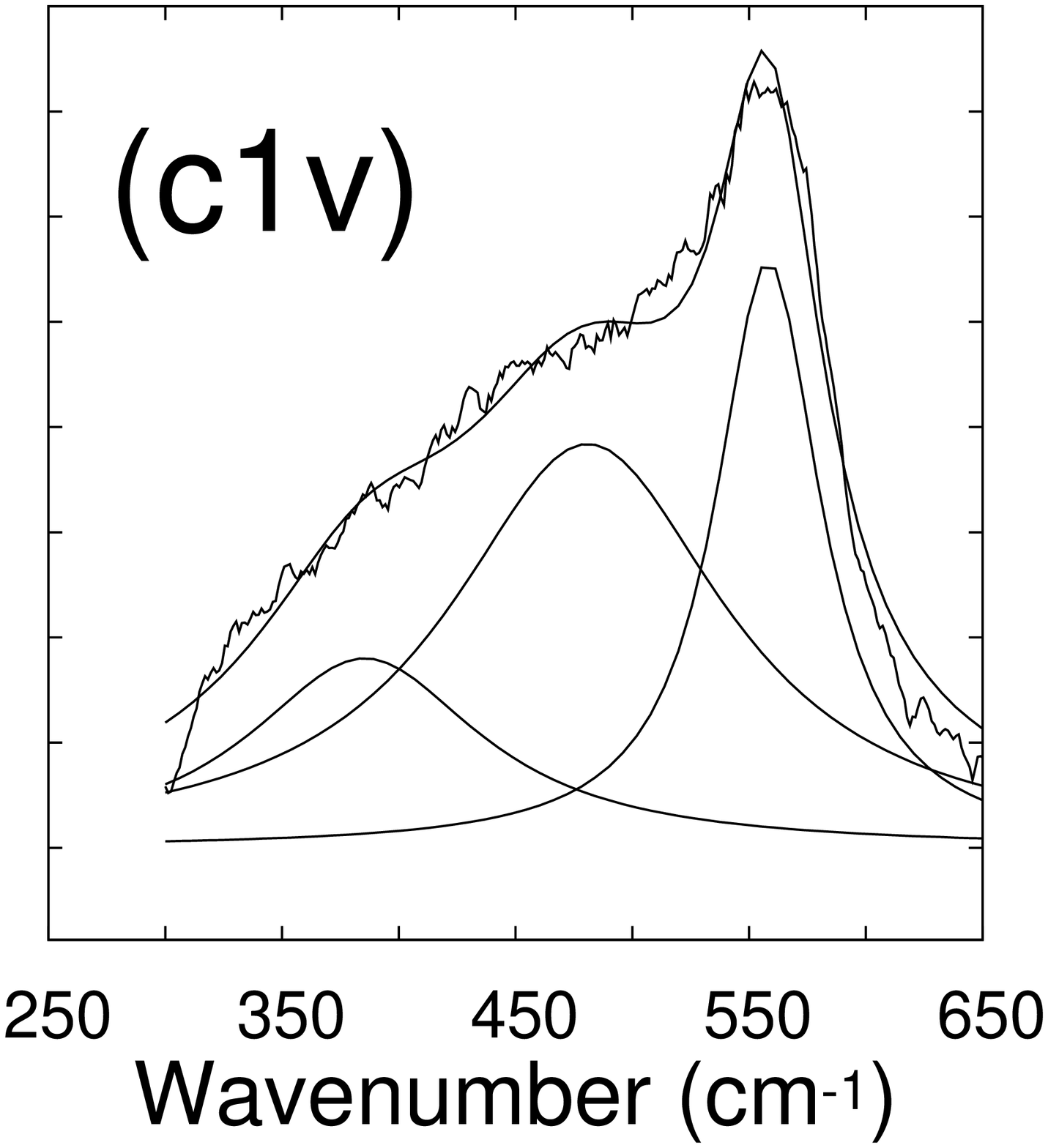, width=0.9in, angle=-90}
\hfil
\epsfig{file=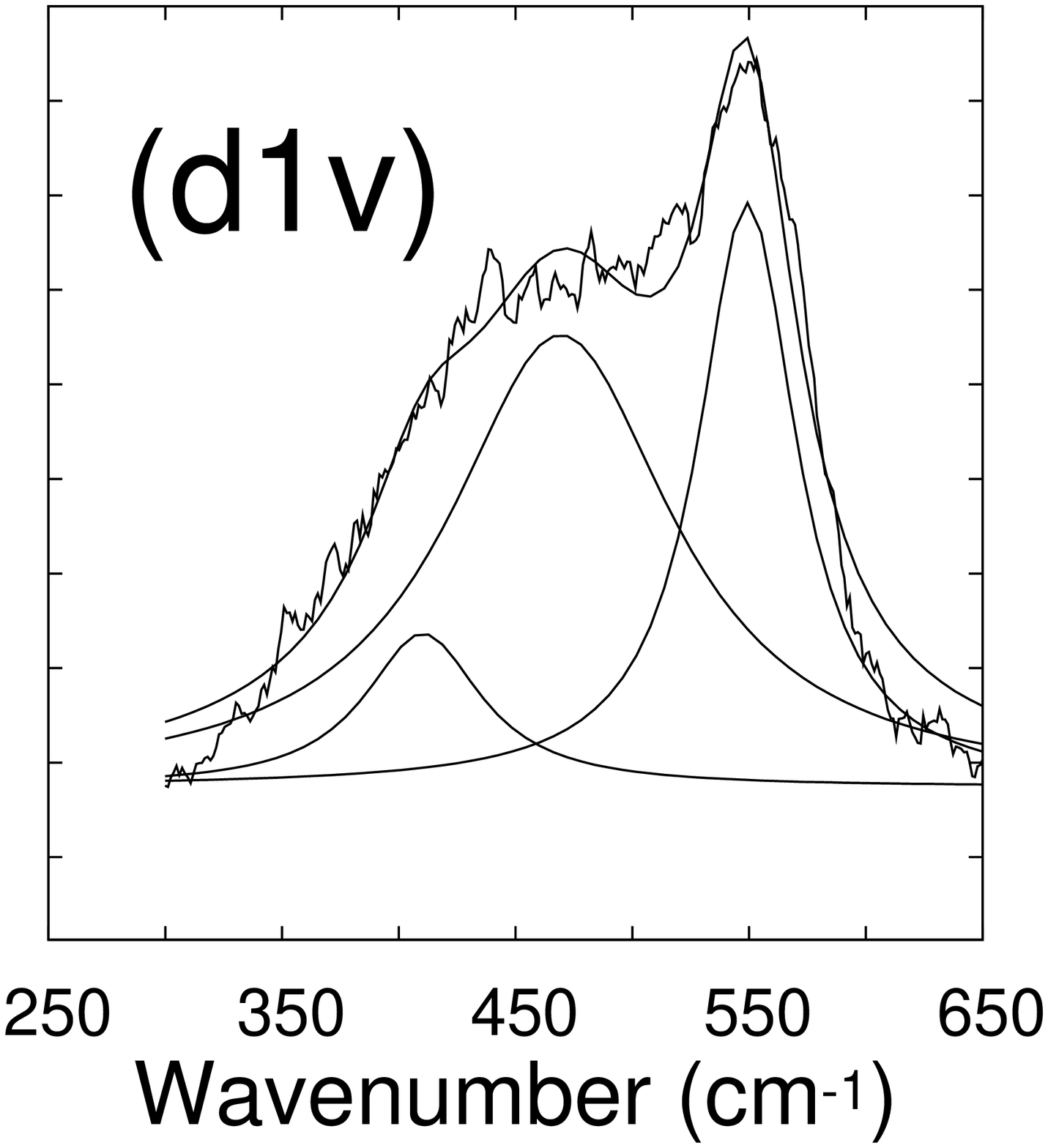, width=0.9in, angle=-90}
\hfil
\epsfig{file=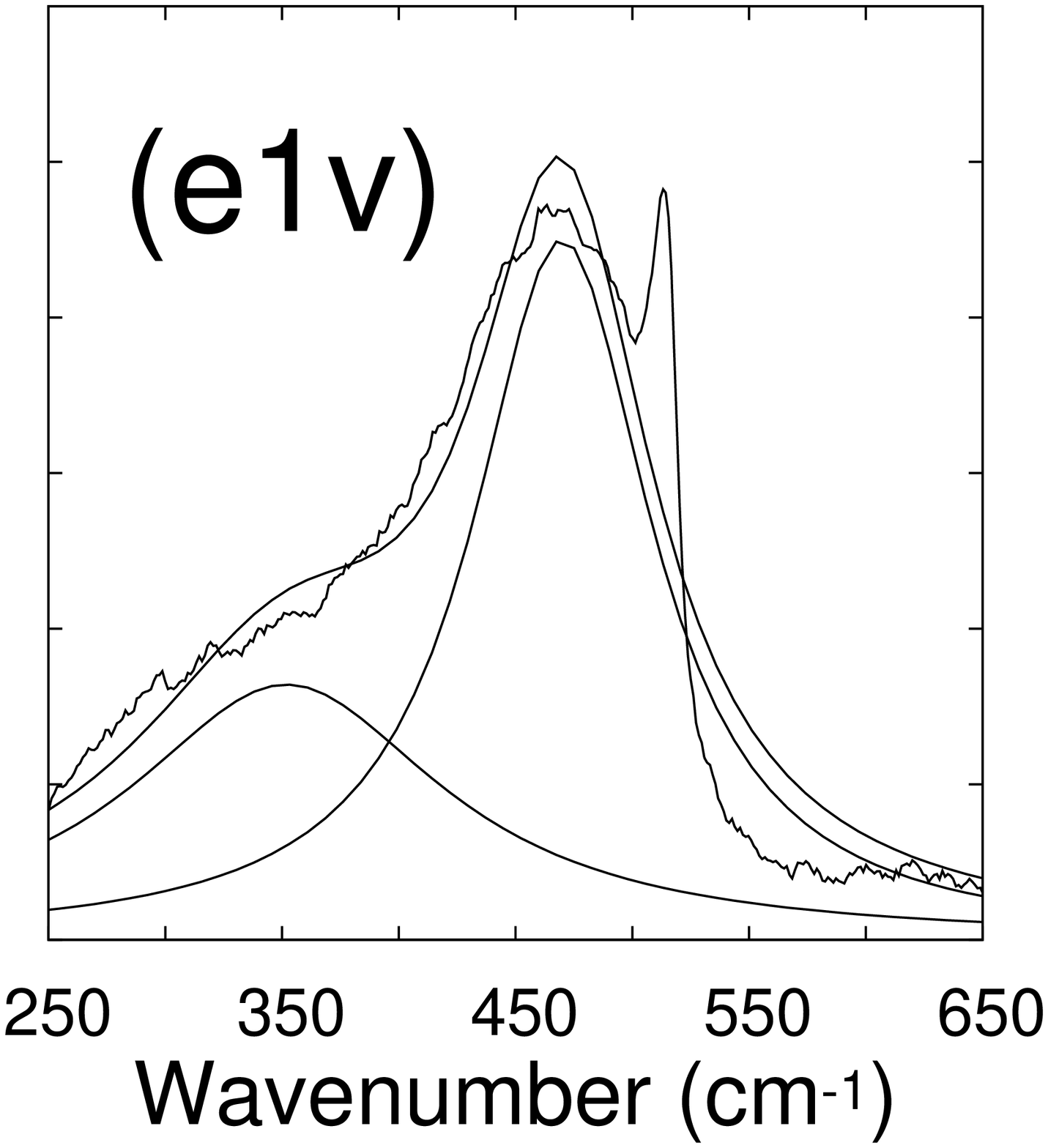, width=0.9in, angle=-90}
\end{center}
\caption{\sl Raman spectra of (A) as grown samples (a1), (b1), (c1), (d1), (e1) and (B) vacuum annealed samples (a1v), (b1v), (c1v), (d1v) and (e1v). Also seen are the deconvoluted peaks assigned to amorphous silicon ($\rm 310cm^{-1}$), wutzite structure of ZnO ($\rm 438cm^{-1}$) and with defect related peak of ZnO ($\rm 570cm^{-1}$).}
\end{figure}

\begin{figure}[h]
\begin{center}
\epsfig{file=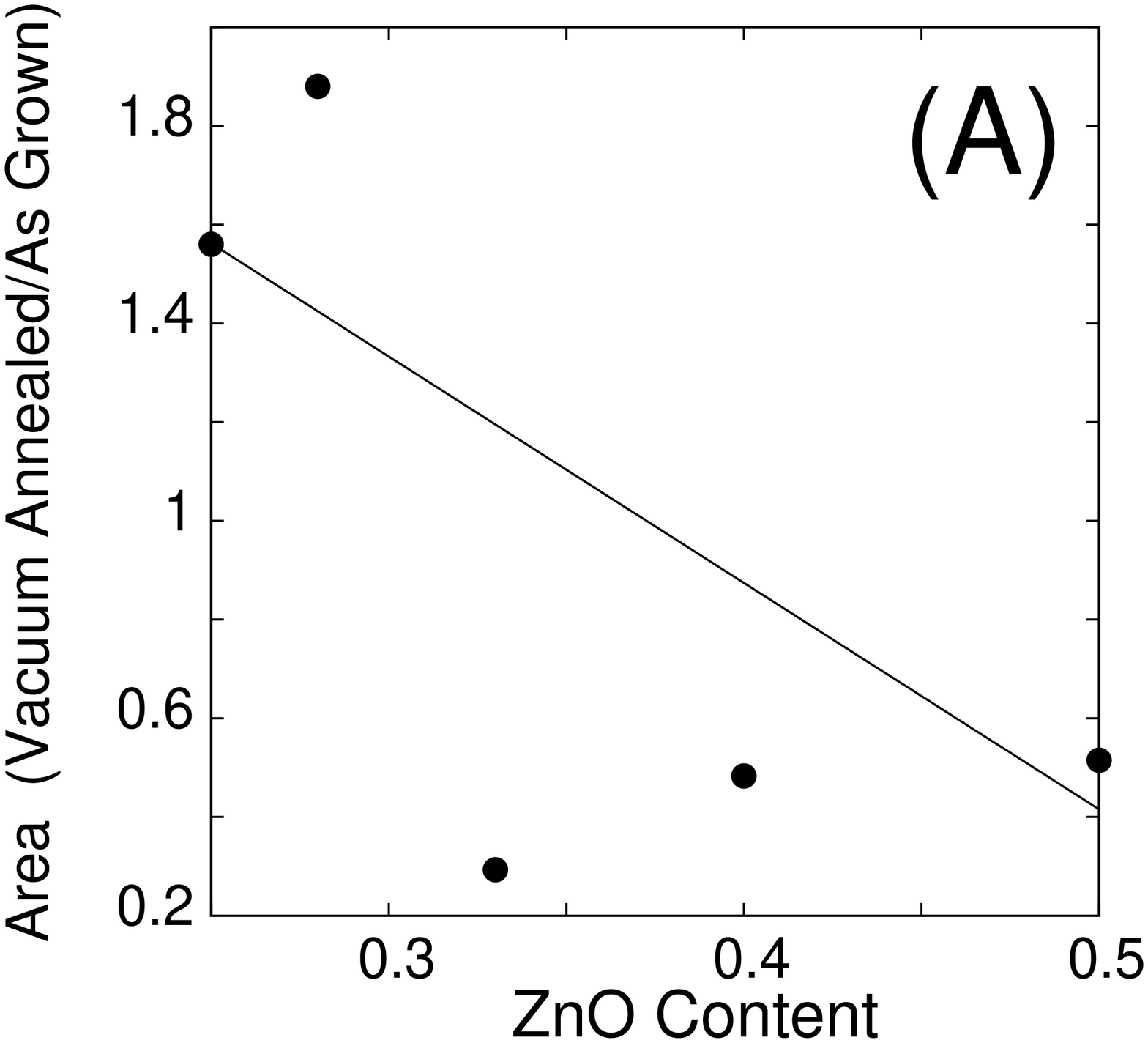,
width=1.5in, angle=-90}
\hfil
\epsfig{file=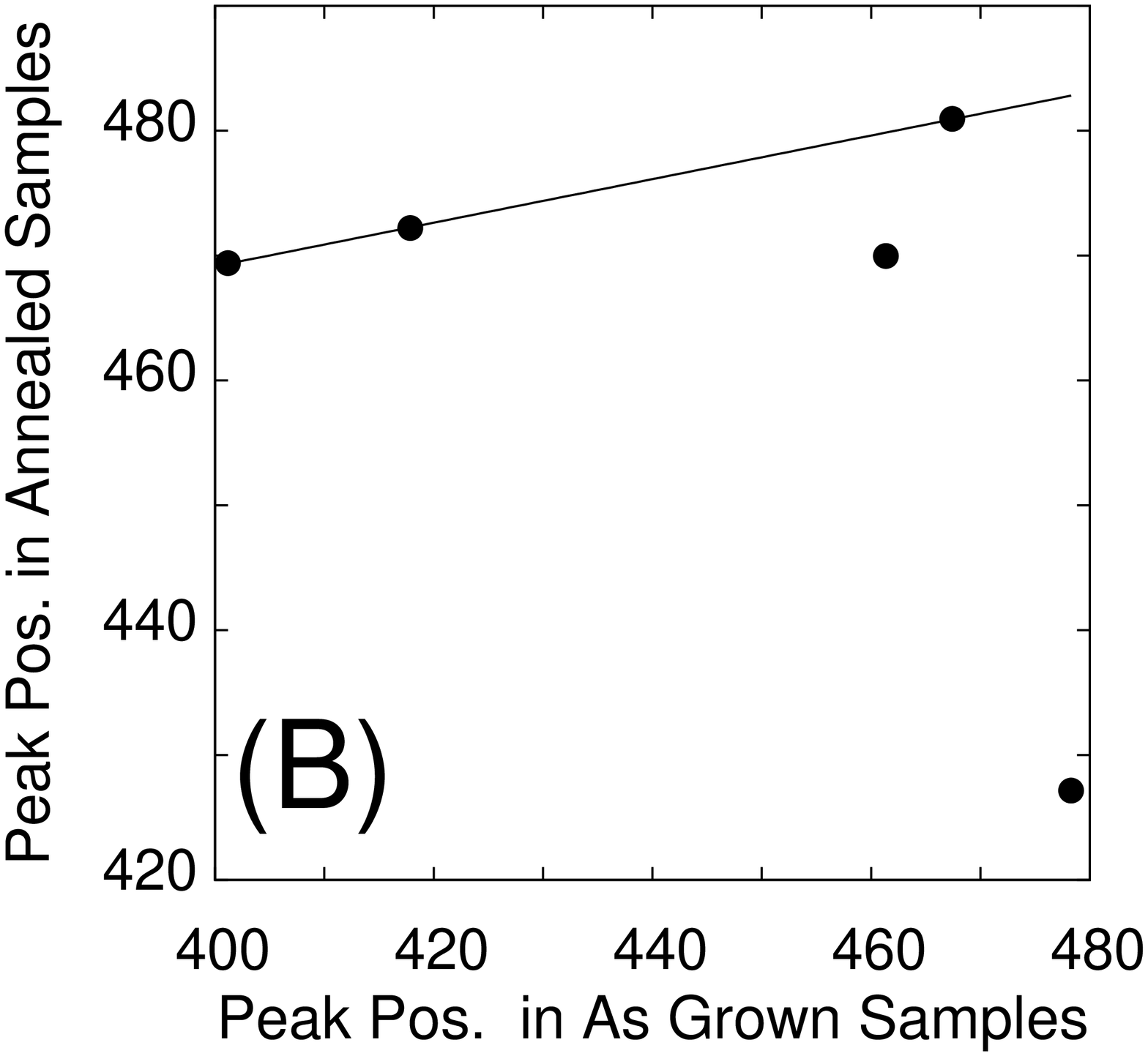, width=1.5in, angle=-90}
\hfil
\epsfig{file=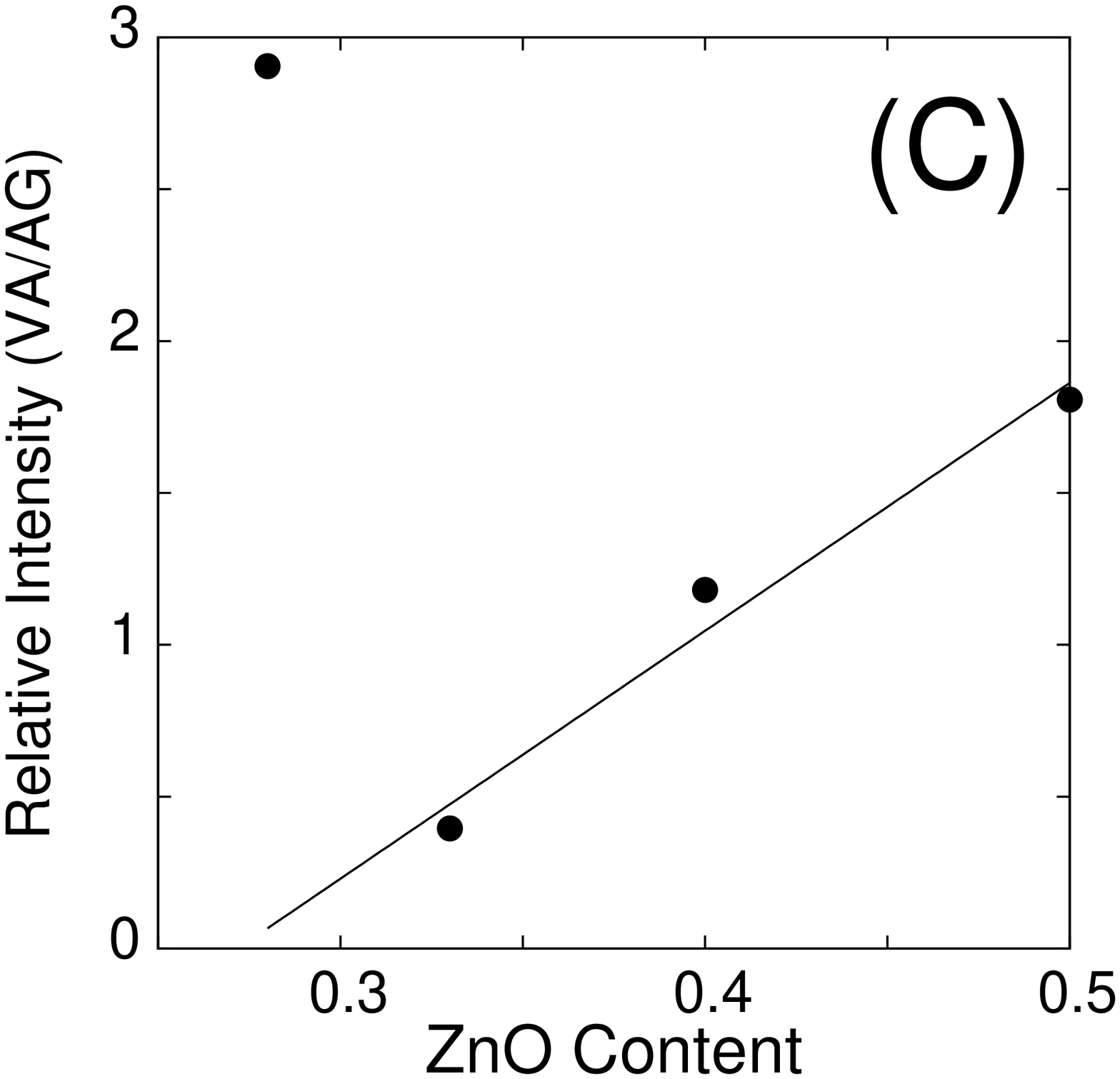,
width=1.5in, angle=-90}
\end{center}
\caption{\sl (A) Relative change in Intensity of 438 $cm^{-1}$ peak in vacuum annealed samples to that in as grown samples with respect to zno content, (B) Variation of peak position in the 438 $cm^{-1}$ peak in vacuum annealed samples with respect to that in the as grown samples and (C) Relative change in Intensity of 560 $cm^{-1}$ peak in vacuum annealed samples to that in as grown samples with respect to ZnO content.}
\end{figure}

\begin{figure}[h]
\begin{center}
\epsfig{file=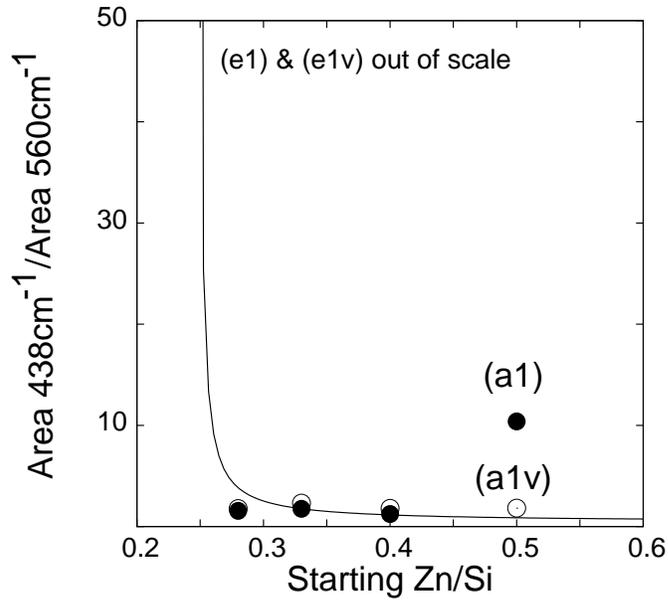, width=3in, angle=-90}
\end{center}
\caption{\sl Relative presence of defedt related ZnO to wurtzite ZnO (Area 438$cm^{-1}$/Area 565$cm^{-1}$ from Raman Spectra) for varying ZnO content in film for (A) vacuum annealed (B)as deposited films.}
\end{figure}

\begin{figure}[h]
\begin{center}
\epsfig{file=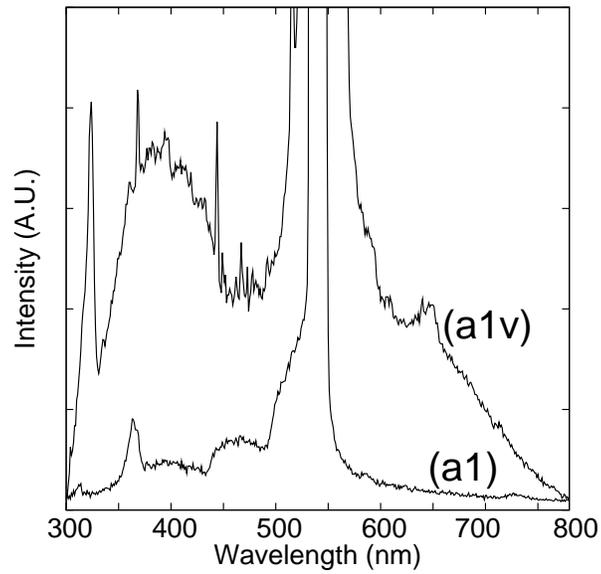, width=3in, angle=-90}
\end{center}
\caption{\sl PL spectra of samples (a1) and (a1v). (Counts of (a1) have been scaled by 3 (i.e.X3) to compare the spectra.)}
\end{figure}

\begin{figure}[h]
\begin{center}
\epsfig{file=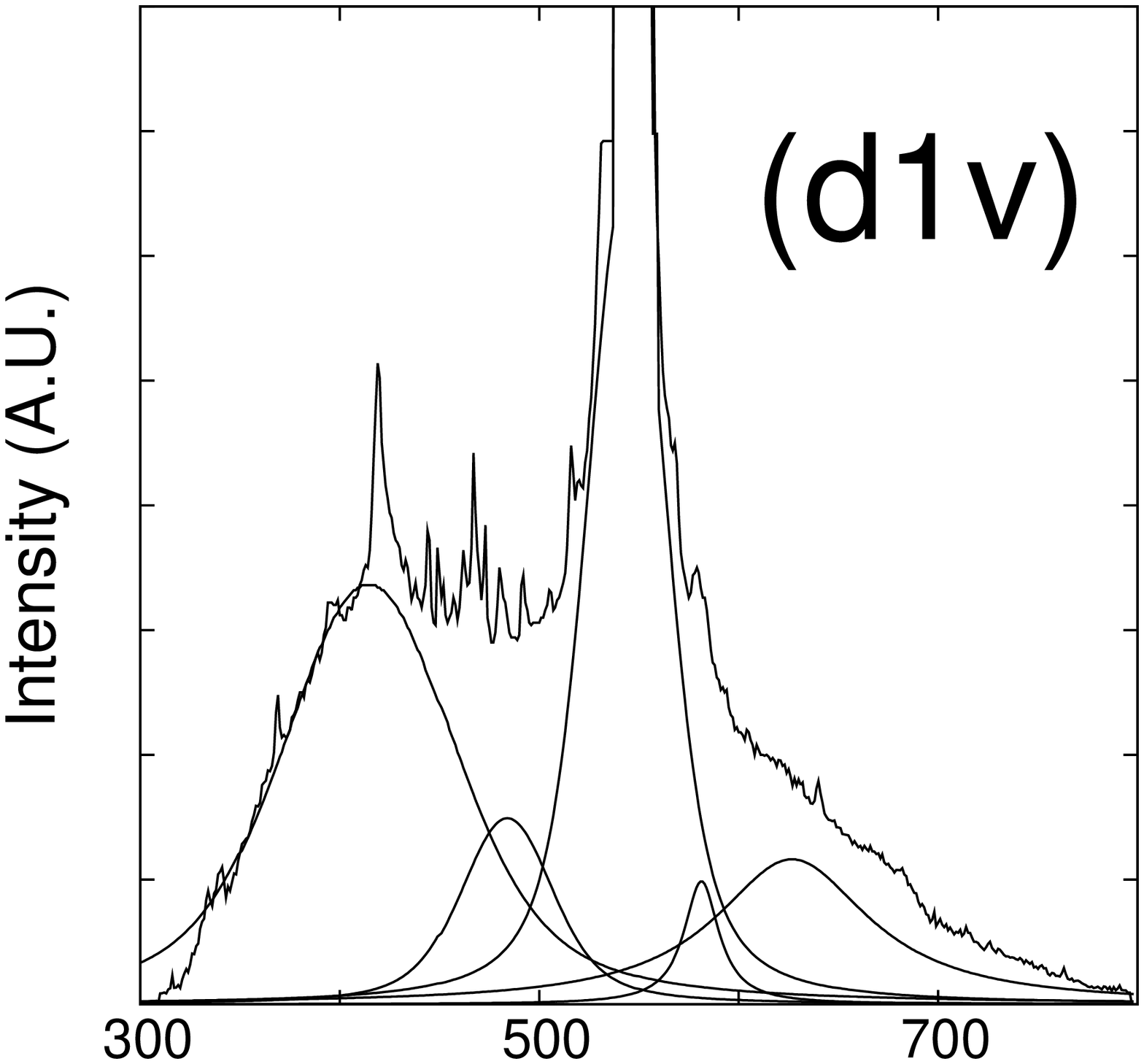, width=2.15in, angle=-90}
\vfil
\epsfig{file=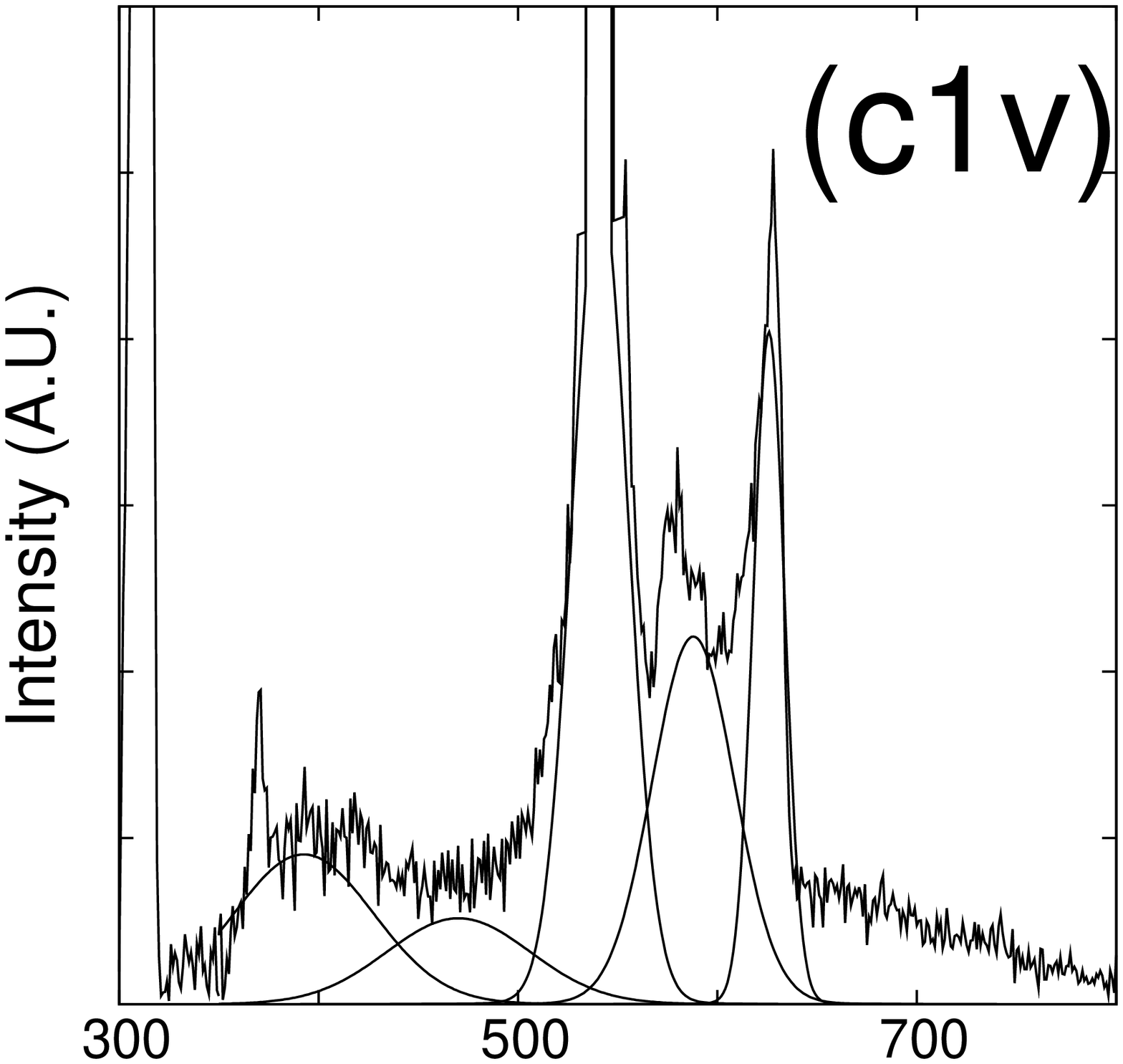, width=2.15in, angle=-90}
\vfil
\epsfig{file=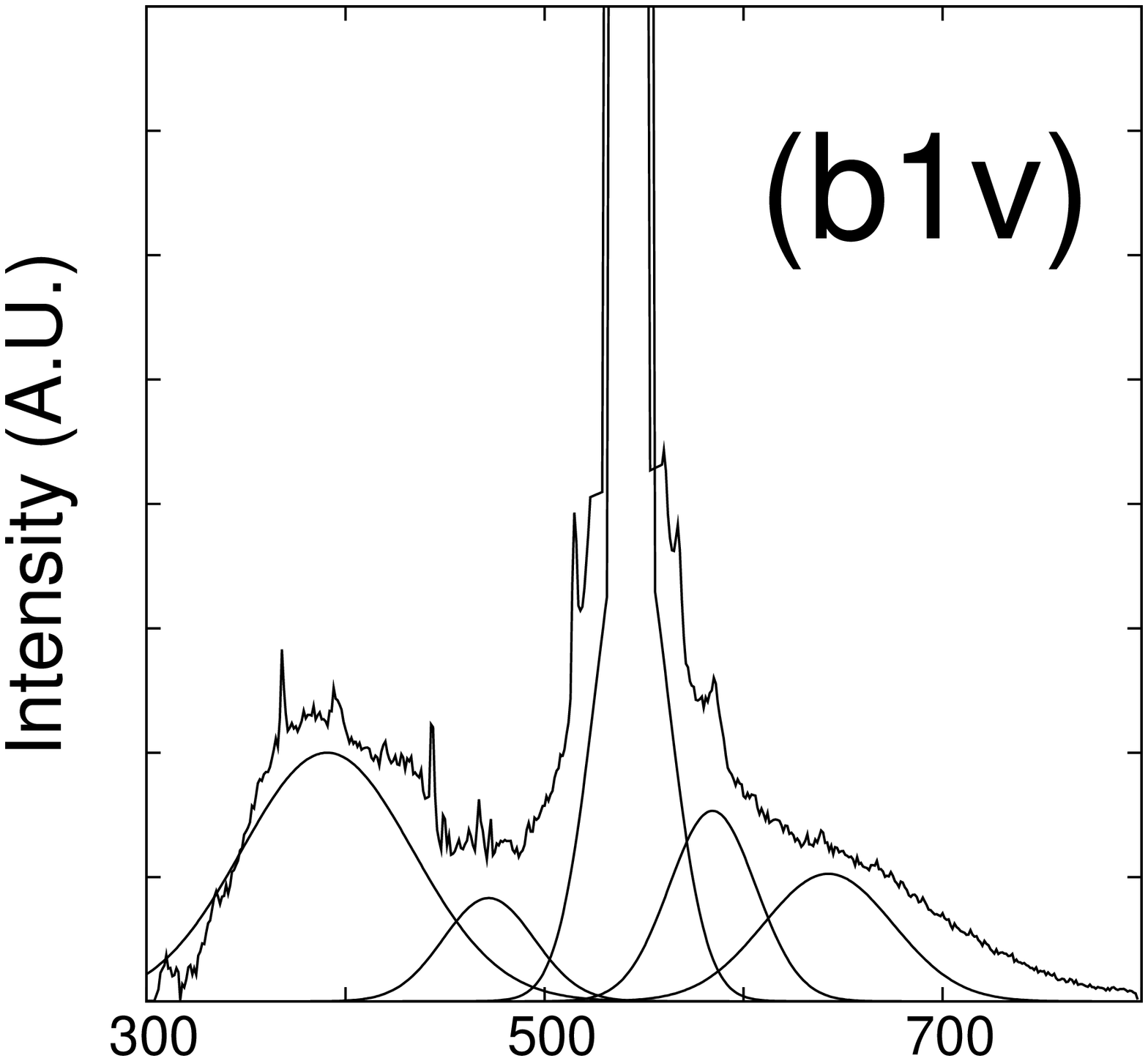, width=2.15in, angle=-90}
\vfil
\epsfig{file=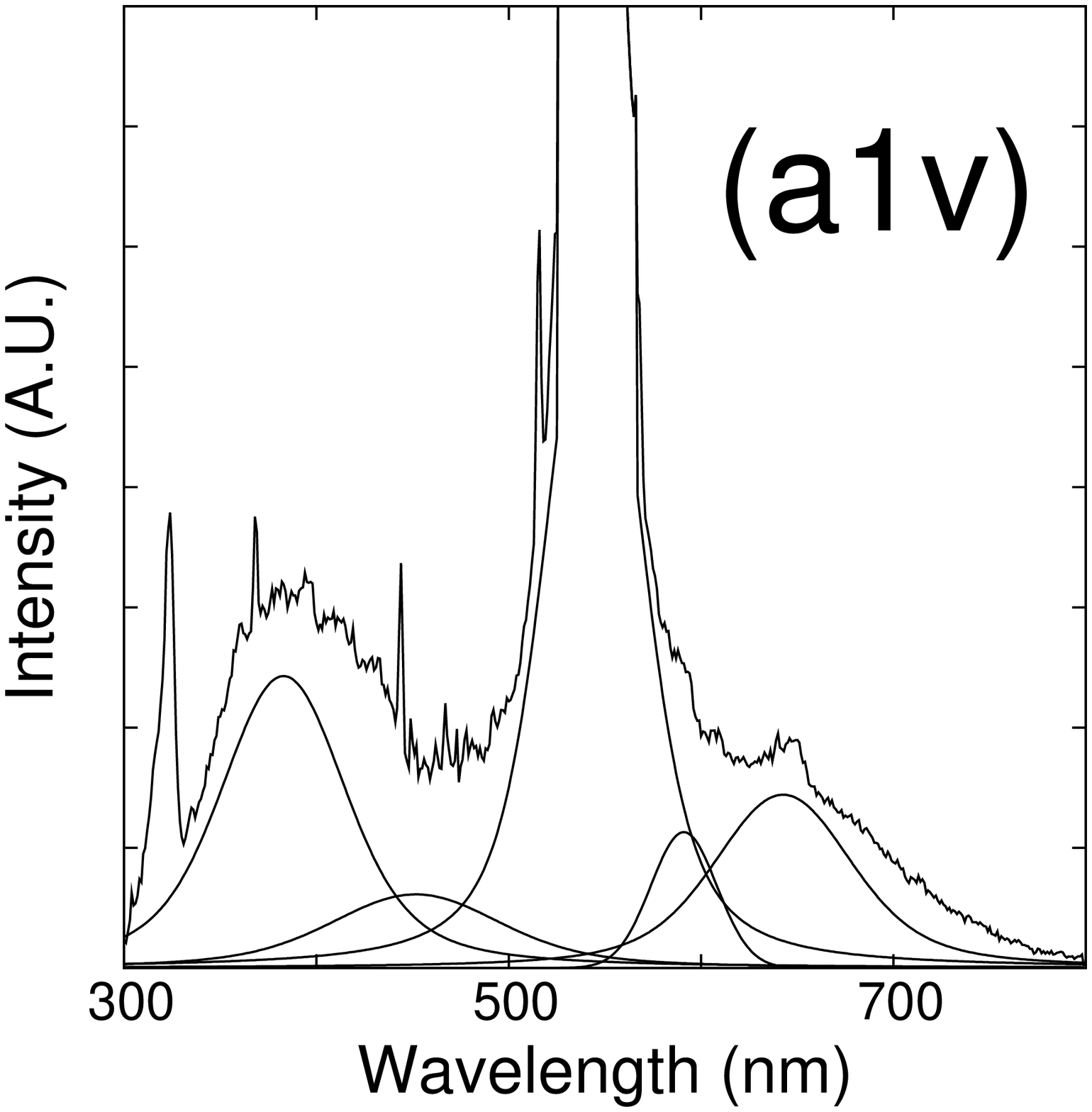, width=2.15in, angle=-90}
\end{center}
\caption{\sl PL of samples (a1v), (b1v), (c1v) and (d1v).}
\end{figure}

\begin{figure}[h]
\begin{center}
\epsfig{file=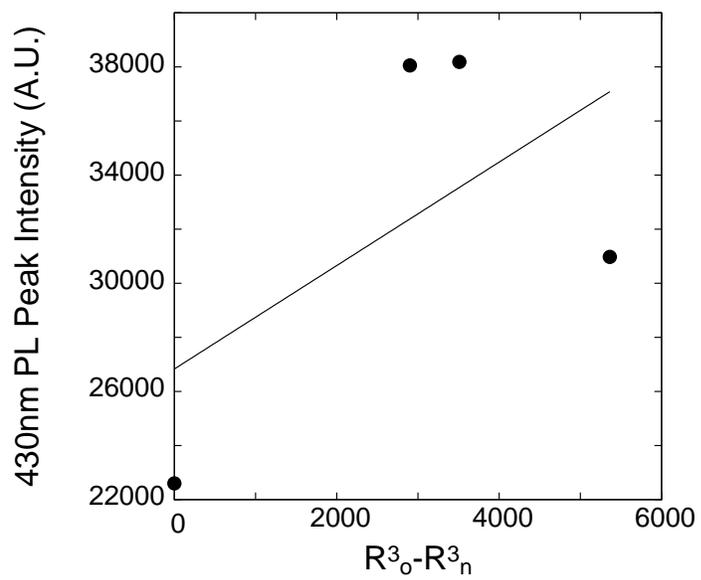, width=3in, angle=-90}
\end{center}
\caption{\sl Plot of shell volume with respect to intensity of 430nm observed in photoluminescence (method described in the text).}
\end{figure}

\end{document}